 \documentclass[final,5p,times,twocolumn]{elsarticle}

\usepackage{amssymb}

\usepackage[fleqn]{amsmath}
\usepackage{bm}
\usepackage{color}
\usepackage{pdfpages}
\usepackage{mmacells}

\newcommand{\dd}{\mathrm{d}} 
\newcommand{\vett}[1]{\bm{#1}}

\newcommand{\eg}{{\em e.g. }}
\newcommand{\ie}{{\em i.e. }}
\newcommand{\etal}{{\em et al. }}

\newenvironment{myindentpar}[2]%
{\begin{list}{}%
         {\setlength{\leftmargin}{#1}\setlength{\rightmargin}{#2}}%
         \item[]%
}
{\end{list}}

\journal{Current Opinion in Colloid and Interface Science}

\begin{document}

\begin{frontmatter}



\title{Elasticity and Stability of Shape Changing Structures}

 \author[label1]{Douglas P. Holmes}
 \ead{dpholmes@bu.edu}
 \ead[url]{www.bu.edu/moss}
 \address[label1]{Department of Mechanical Engineering, Boston University, Boston, MA, 02215 USA}



\begin{abstract}
As we enter the age of designer matter -- where objects can morph and change shape on command -- what tools do we need to create shape--shifting structures? At the heart of an elastic deformation is the combination of dilation and distortion, or stretching and bending. The competition between the latter can cause elastic instabilities, and over the last fifteen years these instabilities have provided a multitude of ways to prescribe and control shape change. Buckling, wrinkling, folding, creasing, snapping have become mechanisms that when harmoniously combined enable mechanical metamaterials, self--folding origami, ultralight and ultrathin kirigami, and structures that appear to grow from one shape to another. In this review, I aim to connect the fundamentals of elastic instabilities to the advanced functionality currently found within mechanical metamaterials.

\end{abstract}


\end{frontmatter}


\section{Introduction}
\label{Introduction}

How do objects change shape? This question has formed the basis of entire branches of mechanics and physics dating back centuries, and so it is reasonable to wonder what new questions and challenges remain. To drastically change an object's shape, it should possess some degree of {\em softness} -- either in a material sense, \eg having a low elastic modulus, or in a geometric sense, \eg having slenderness. This softness comes at a price -- large deformations introduce nonlinear responses and instabilities. Material nonlinearities are common with traditional engineering materials, {\em i.e.} metals, wood, ceramics, which, whether they are brittle or ductile will tend to irreversibly deform in response to small strains, either by fracturing or flowing plastically. Softer materials, like rubbers, gels, and biological tissues can often withstand moderate amounts of strain without reaching a material limit, and so they can reversibly withstand elastic instabilities without permanent deformation, exhibiting geometric nonlinearities by bending, buckling, wrinkling, creasing, and crumpling. It is perhaps no surprise then, that a resurgence in studying elastic instabilities coincided with the emergence of new, simple, and inexpensive ways to prepare soft elastomers in any desired shape and size, thus enabling researchers to study how instabilities could perform useful functions. Now the study of how to utilize elastic instabilities for mechanical functionality brings together the disciplines of soft matter physics, mechanics, applied mathematics, biology, and materials science with the aim to extend our understanding of structural stability for both generating form and function.

\section{Elasticity of Slender Structures}\label{elasticity}
\subsection{Stretching and Bending}

Slenderness, embodied by the canonical forms: rods, plates, and shells, provides the most direct way to deform a structure, as the reduced dimensionality enables large deformations while the material stress remains low, \ie $\sigma/E \ll 1$, where $\sigma$ is the maximum principal stress and $E$ is Young's elastic modulus. Thin structures are highly susceptible to instability, and this is due in large part to their tendency to deform by bending\footnote{Readers interested in a more thorough understanding of the ideas presented throughout this review should consult the Supplementary Information~\cite{HolmesSI}.}. The fact that these structures are by definition thinner in one dimension than the other two motivated the development of models of elastic deformation of lower spatial dimension, {\em i.e.} reduced order models, to describe slender structures, such as rods, plates, and shells. With a shell being perhaps the most generic of these forms, we can develop some intuition for how thin structures deform by looking at the strain energy of a thin elastic shell. Shell theories have a common structure: a stretching energy $\mathcal{U}_s$, which accounts for extension or compression of the middle surface of the shell and is linear in the shell thickness, $h$, and a bending energy $\mathcal{U}_b$, which accounts for the curvature change of the deformed shell, and is dependent on $h^3$. The strain energy of a thin shell will scale as~\cite{HolmesSI}
\begin{equation}
\label{U}
\mathcal{U} \sim  \underbrace{Y \int \varepsilon^2 \ \dd \omega}_{\mathcal{U}_s} + \underbrace{B \int \kappa^2 \ \dd \omega}_{\mathcal{U}_b}
\end{equation}
\begin{figure*}[t]
\begin{center} 
\includegraphics[width=2\columnwidth]{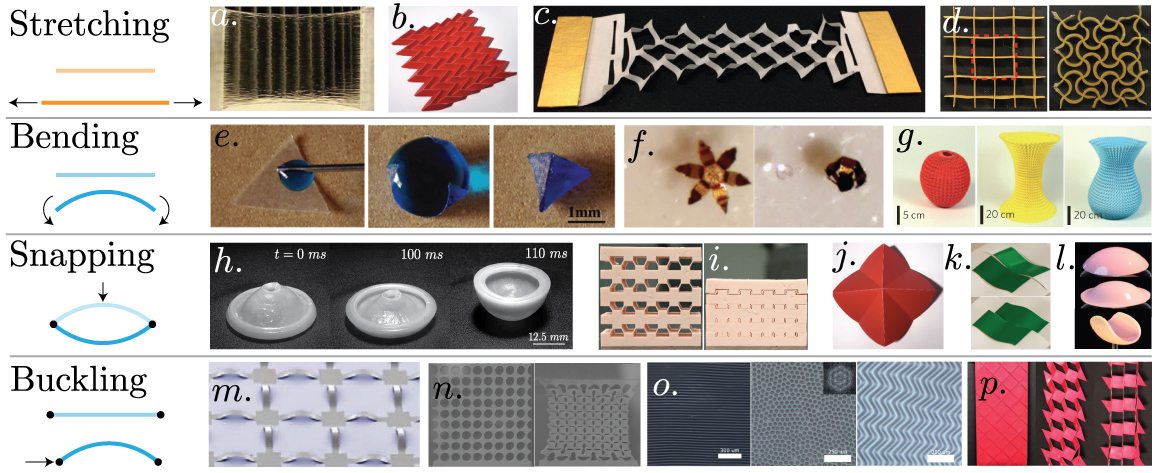} 
\end{center}
\vspace{-7mm} 
\caption{\textsc{Stretching:} $a.$ Highly stretchable gel using double--network gels~\cite{Sun2012}, $b.$ Miura--ori fold~\cite{Brunck2016}, $c.$ extremely stretchable thin sheets via kirigami~\cite{Blees2015}, and $d.$ negative swelling gels~\cite{Liu2016}.  \textsc{Bending:} $e.$ capillary origami~\cite{Py2007}, $f.$ tetherless microgrippers~\cite{Leong2009}, and $g.$ programming curvature with origami~\cite{Dudte2016}. \textsc{Snapping:} $h.$ snapping poppers~\cite{Pandey2014}, $i.$ snapping elements that absorb elastic strain energy~\cite{Shan2015}, $j.$ bistable water bomb fold~\cite{Brunck2016}, $k.$ bistable kirigami unit cells~\cite{Yang2018}, and $l.$ swelling--induced snapping shells~\cite{Pezzulla2018}. \textsc{Buckling:} $m$. buckled silicon membranes~\cite{Rogers2010} $n.$ negative Poisson ratio~\cite{Bertoldi2010}, $o.$ wrinkle patterns~\cite{Cai2011}, $p.$ buckling--induced kirigami~\cite{Ahmadkatia2017} \label{fig-overview}} 
\end{figure*}
where $Y=Eh/(1-\nu^2)$ is the stretching rigidity, $B=Eh^3/[12(1-\nu^2)]$ is the bending rigidity, $\nu$ is Poisson's ratio, and $\dd \omega$ is the area element. To gain some physical intuition about a particular problem it is often adequate to simply consider how the relevant energies scale. For instance, by simply comparing the stretching and bending energies, we see that $\mathcal{U}_b/\mathcal{U}_s\sim h^2 (\kappa/\varepsilon)^2$, where $\kappa$ is the average curvature of induced by bending the shell (units of reciprocal length), and $\varepsilon$ is the average strain induced when stretching the shell (unitless). Since the shell is thin, this quantity must be very, very small, indicating that it is far easier to bend a thin structure than it is to stretch it. This insight helps explain why thin structures are prone to instability: if you try to shorten the length of a thin rod by compressing it, the rod would much rather bend than be compressed, and to bend it must buckle.

The remainder of this review deals with the challenge of controlling the shape change of a structure while overcoming the constraints on bending and stretching it. An overview of how researchers have overcome these constraints is presented in Fig~\ref{fig-overview}, which highlights how tailoring materials, geometry, and topology can enable structures to stretch and bend in unconventional ways, and how elastic instabilities enable structural morphing and metamaterial behaviors such as negative Poisson ratio and negative swelling. By building upon the fundamentals of elasticity (Section~\ref{elasticity}), and harnessing elastic instabilities for enhanced functionality (Section~\ref{stability}), we are now ushering in an age of {\em programmable matter} (Section~\ref{programmable}). The multitude of approaches for changing an object's shape share similar techniques -- mechanical metamaterials are built around buckling and snapping mechanisms, {\em origami} and {\em kirigami} create local regions that stretch easily, shape--shifting structures swell more or less locally -- and the purpose of this review is to provide the reader the insight to see the underlying principles that govern shape change.

\subsection{Scaling}

Using the scaling relations of a thin structure's stretching and bending energy, we were able to quickly see why thin object bend rather than stretch. This approach can be useful for understanding the relevant physics in a whole range of phenomena, and we will review some of the most relevant examples here. 

\subsubsection{Elastogravity Length}

Let's consider a simple question: if I slide a sheet of paper over the edge of my desk, at what length will this sheet begin to bend under its own weight? Gravity is the relevant force here, and we know that the gravitational potential energy scales as
\begin{equation}
\label{Ug}
\mathcal{U}_g \sim \rho g \int \delta^2 \ \dd \omega,
\end{equation}
where $\rho$ is the volumetric density of the paper, $g$ is the acceleration due to gravity, and $\delta$ is the distance over which gravity is acting. To determine when gravity will bend a thin sheet, we need to consider the term $\kappa$ in equation~\ref{U} more carefully. Here, bending represents a vertical deflection $\delta$ occurring relative to our unknown length $\ell$. Since $\kappa$ is a curvature, for small deflections we may write $\kappa \approx  \partial^2 \delta /\partial x^2$, such that from dimensional analysis we may write $\kappa \sim \delta/\ell^2$. We expect the sheet of paper will bend when the energetic potential from gravity is on the same order as the energetic resistance to bending, and so by equating equation~\ref{Ug} and $\mathcal{U}_b$ we arrive at a characteristic {\em elastogravity} length scale\footnote{Depending on the choice of $B$ (which often takes the form of $B\sim Er^4$ for a rod) and the choice of $\rho$ (which may be taken to be an areal density), this exponent may change accordingly.}\cite{Foppl1907}
\begin{equation}
\label{Leg}
\ell_{eg} \sim \left(\frac{B}{\rho g}\right)^{1/4}.
\end{equation}
Typical office paper has a bending rigidity of $B=40$ mN$\cdot$m, and density of $\rho=800$ kg/m$^3$, meaning that an overhang of about 47 mm (1.85 in) will cause the paper to start to sag (Fig.~\ref{fig-sect1}a). Beyond being a useful back--of--the--envelope calculation, the {\em elastogravity} length has been shown to play an important role in determining the wavelength of wrinkling soap films~\cite{Milner1989}, the ability of thin elastic sheets to ``grab'' water~\cite{Reis2010}, the curling~\cite{Callan2012} and twisting~\cite{Lazarus2013} of an elastic rod, the viscous peeling of plates~\cite{Lister2013}, and the buckling~\cite{Bense2017} and the wrinkling~\cite{Pineirua2013} of elastic sheets on water.

This balance of bending and gravity is also what determines the length of curly hair~\cite{audoly10, Miller2014}. If the curl is treated as a circular spring that is opening under the force of gravity, the bending stiffness of the hair will be $k_\text{curl}\sim EI \kappa_n^3$, where $\kappa_n$ is the curvature of the curl, or the intrinsic, natural curvature of the hair. Gravity acts upon each curl by $\rho g A L$, where $\rho$ is the density (mass/volume), $A$ is the cross sectional area of the hair, and $L$ is the total length of the hair. Each individual curl, or circular spring, will open by an amount $ z_\text{curl} \sim \rho g A L/ k_\text{curl}$, and if there are $n$ curls, such that $n\sim L \kappa_n$, the vertical displacement of the free end of the hair should be~\cite{audoly10}
\begin{equation}
z \sim \frac{\rho g A L^2}{EI \kappa_n^2}.
\end{equation}

A similar argument can be used to design a Slinky. A Slinky is little more than a very floppy spring, and one of the most iconic features of this toy is that the can be bent into an arch. The bending energy for this discrete structure is slightly different than the continuous form given by $\mathcal{U}_b$~\cite{Holmes2014}, however the concept is the same. Balancing the spring's effective bending rigidity against the gravitational potential yields indicates that this Slinky would need $71$ rings to form a stable arch~\cite{Holmes2014}\footnote{Using $n_r \sim \overline{EI}/(mgRh)$, with radius $R=34.18$mm, thickness  $h=0.67$mm, mass per ring $m=2.49$g, and an experimentally measured effective bending rigidity of $\overline{EI}=40 \times 10^{-6}$ N $\cdot$ m$^2$.}. Since the Slinky typically has about 82 rings, cut off about 11 or more and it won't be as fun to play with.~\cite{Holmes2014}



\subsubsection{Elastocapillary Length}

We are familiar with fluids deform thin structures through inertia, for example, consider the flapping of a flag in the wind~\cite{Argentina2005}. In contrast, capillary forces are typically negligible at macroscopic scales. However, at micro and millimetric scales, surface tension can cause thin structures to bend. So, a natural question to ask is: at what length scale should I start to worry that capillary forces will deform my structure? Here, we are concerned with surface energy, which scales as
\begin{equation}
\label{Ugamma}
\mathcal{U}_\gamma \sim \gamma \int \dd \omega,
\end{equation}
where $\gamma$ is the surface tension of the liquid. The most elegant scenario to imagine was outlined by Roman and Bico in their review papers on this topic~\cite{Roman2010, Bico2018} in which they consider a thin strip spontaneously wrapping around a wet cylinder of radius $R$. For the strip to wrap around the cylinder, it must adopt a curvature of $\kappa \sim R^{-1}$. Here again we can balance bending energy $\mathcal{U}_b$ with surface energy (eq.~\ref{Ugamma}) to arrive at a characteristic {\em elastocapillary} length~\cite{Bico2004, Py2007} 
\begin{equation}
\label{Lec}
\ell_{ec} \sim \left(\frac{B}{\gamma}\right)^{1/2}.
\end{equation}
With this simple example, the prefactor can be worked out to be exactly $1/2$~\cite{Roman2010}, indicating that a sheet of office paper will spontaneously wrap around a cylinder wet with water ($\gamma =72$ mN$\cdot$m) as long as its radius is greater than 372 mm (Fig.~\ref{fig-sect1}b). Intuitively, this makes sense -- paper is intrinsically flat, and so it should conform to a cylinder regardless of $\gamma$ as $R\rightarrow\infty$, but below some critical $R$ bending will be too costly. The role of surface stresses is a highly active research area at the moment~\cite{Zhao2018}, and those interested in this brief primer would be better served reading the work of Bico and Roman~\cite{Roman2010, Bico2018}, along with recent reviews on deforming soft solids~\cite{Style2017}, bundling fiber arrays~\cite{Duprat2015}, and approaches for using fluids to assemble structures~\cite{Crane2013}.

\subsubsection{Warping Wafers}

It is fair to say that one of the most classical examples of a shape changing structure is the bending of a bimetallic strip when it is heated~\cite{Timoshenko1925}. 
While thin bimetallic strips and beams bend uniformly when subjected to a change in temperature, thin plates do not. This is perhaps familiar to those who have cooked in the oven with a metallic baking sheet, as it may buckle and warp when heated above a certain temperature. The bowing and warping of thin plates was a particularly important problem relating to the deposition of thin metallic films onto silicon wafers~\cite{Stoney1909, Freund1999}. Homogenous heating of a bimetal plate will endow the plate with a homogenous natural curvature $\kappa_n$, causing it to bend into the segment of a spherical cap, which has a positive Gaussian curvature. Gauss's famous {\em Theorema Egregium} states that you cannot change the intrinsic curvature of a surface without stretching it, and so this deformation comes at the cost of stretching the plate's middle surface. Eventually, the energetic cost for the plate to bend into a cylinder becomes lower than the cost to continue bending into a spherical cap, and so the bowing wafer will buckle into a cylindrical shape~\cite{HolmesSI, Seffen2007}.

\begin{figure} 
\begin{center} 
\includegraphics[width=1\columnwidth]{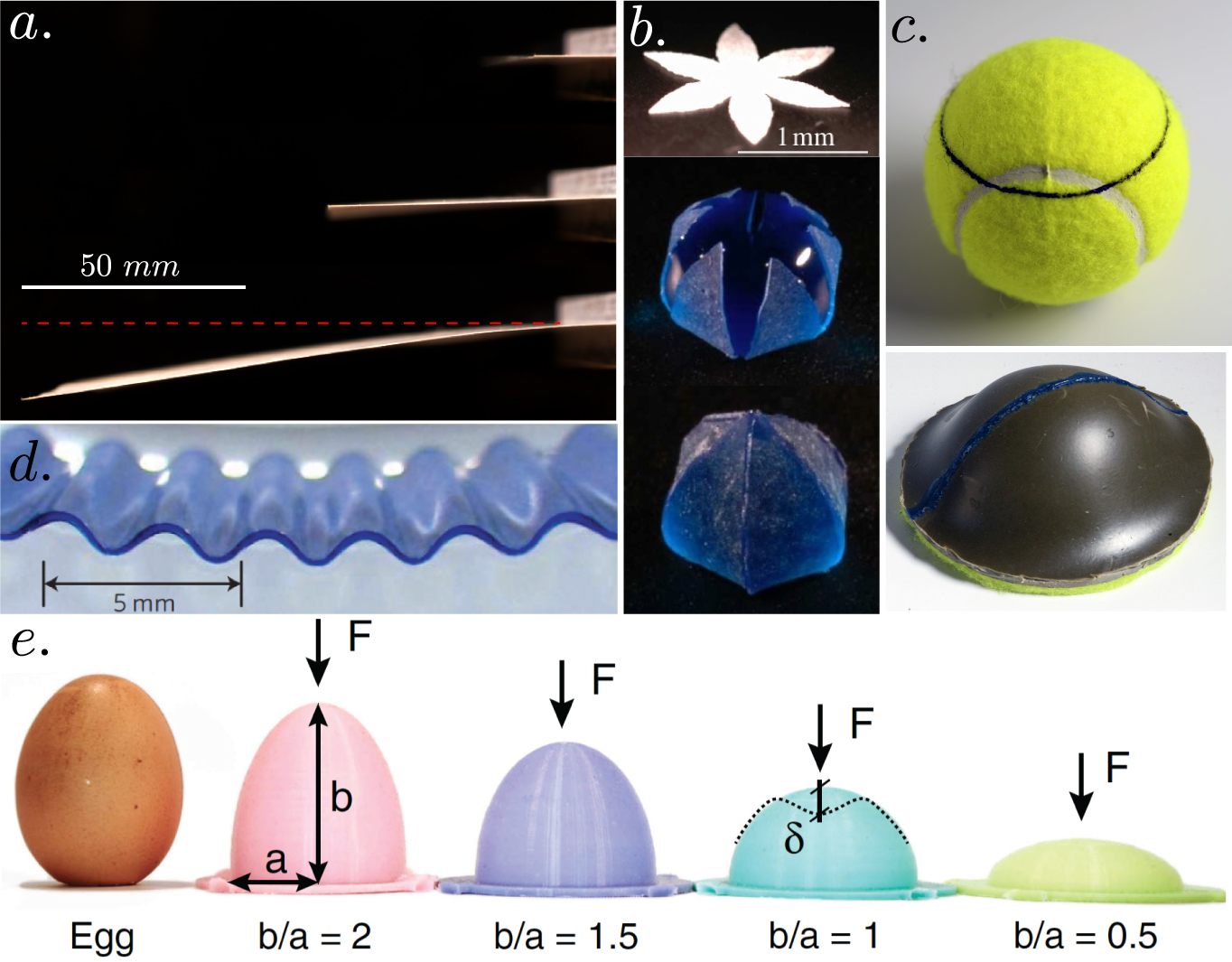} 
\end{center}
\vspace{-7mm} 
\caption{$a.$ Demonstration of the elastogravity length with a sheet of paper. $b.$ Elastocapillary encapsulation adapted from~\cite{Py2007}, $c.$ Snap--through eversion of a spherical cap, demonstrated with a tennis ball, adapted from~\cite{Taffetani2018}, $d.$ Wrinkles on a soft substrate adapted from~\cite{Brau2011}, and $e.$ a demonstration of geometric rigidity~\cite{lazarus2012geometry}. \label{fig-sect1}} 
\end{figure}

The cost of stretching the plate's middle surface into a spherical cap is quantified by $\varepsilon$ in equation~\ref{U}, and this will scale as the square of the natural curvature $\kappa_n$ times the plate radius $r$, such that $\varepsilon \sim (\ell \kappa_n)^2$~\cite{Pezzulla2016}. Alternatively, to avoid stretching the disk will need to maintain its zero
Gaussian curvature, which it can accomplish by bending into a cylinder. If it deforms into a cylindrical shape, the stretching energy is zero while the sheet has to suppress the curvature along one direction, such that the bending energy scales as it is written in equation~\ref{U}, with $\kappa \sim \kappa_n$~\cite{Pezzulla2016}. By balancing the bending and stretching energies from equation~\ref{U}, we find that the plate should buckle when~\cite{Pezzulla2016}
\begin{equation}
\label{natCurv}
\kappa_n \sim \frac{h}{\ell^2},
\end{equation}
where, for thermal problems the natural curvature can be related to the temperature by calculating the curvature of a beam of given modulus and thickness ratios using Timoshenko's well known result~\cite{Timoshenko1925}. In addition, the prefactor of this equation has been worked out exactly for most plate geometries~\cite{Pezzulla2016}. This scaling analysis also gives rise to a characteristic length scale that is relevant to the anticlastic curvature of bent plates~\cite{Bellow1965, Pomeroy1970}, and boundary layers in thin shells. By simply inverting the relation, we find
\begin{equation}
\label{Lbl}
\ell_{bl}\sim\sqrt{\frac{h}{k_n}} \sim\sqrt{hR},
\end{equation}
where $R$ is the radius of curvature of a given shell. This characteristic length is easy to identify in thin shells. Take a tennis ball, cut it in half, and turn it inside out (Fig~\ref{fig-sect1}c). The flat lip that forms along the boundary of the everted shell has a length on the order of $\ell_{bl}$.

\subsubsection{Snapping Shells}

In the previous section, I encouraged an experiment that involving cutting a tennis ball and turning it inside out. A natural question may be: beginning at the ball's apex, how far down, {\em i.e.} at what latitude from the north pole, should you cut through it to ensure that it can be turned inside out? Turning a shell inside out requires the middle surface of the shell to stretch as it is everted. Pushing on a shell of radius $R$, like a ping pong ball or water bottle, to a depth $\delta$ cause a dent with a characteristic length of $\ell_{bl}\sim \sqrt{\delta R}$ -- the same scaling that appears in equation~\ref{Lbl} -- and this dent is equivalent to a locally everted segment of the shell. This gives us a measure of the stretching strain, with $\varepsilon \sim \delta^2/\ell_{bl}^2\sim\delta/R$~\cite{Taffetani2018}. Deflection of the apex of the shell can be estimated in terms of the geometry of the shell, with $\delta=R(1-\cos \alpha)\sim\alpha^2R$, where $\alpha$ is the planar angle subtended between the pole and the free edge of the shell~\cite{Taffetani2018}. Now the stretching energy of the shell can be found from equation~\ref{U} using $\varepsilon\sim \alpha^2$. To turn a shell inside out, it will adopt a new radius of curvature that is quite close to its original radius of curvature, and the main distinction is that material points that were originally on the outer (inner) surface of the shell are now being compressed (stretched). This comes at the cost of bending the shell, which according to equation~\ref{U} can be found using $\kappa \sim 1/R$. Balancing these two energies, and, for historical reasons, taking the fourth root of the result, gives rise to a dimensionless parameter that characterizes the shell~\cite{Brinkmeyer2012, Taffetani2018}
\begin{equation}
\label{LambdaShell}
\Lambda \sim [12(1-\nu)]^{1/4}\left(\frac{R}{h}\right)^{1/2}\alpha.
\end{equation}
Shells with $\Lambda \gtrsim 5.75$ can be turned inside out, or everted, and remain that way -- {\em i.e.} they are bistable~\cite{Taffetani2018}. For a typical tennis ball, with $R=33.15$mm, $h=3.3$mm, and Poisson ratio of $\nu=1/2$, that means cutting the ball at a height of 15mm from its north pole will yield a bistable shell.

\subsubsection{Wrinkles}

Probably no instability is more responsible for the surge in research interest on the topic of this review over the last fifteen years than wrinkling~\cite{Cerda2002, Cerda2003, Efimenko2005, Huang2007}. Like the others we have encountered so far, this problem also has both a long history~\cite{Allen1969} and a myriad of potential utility~\cite{Li2012}. It is quite easy to see the pattern characteristic of wrinkling by simply compressing the skin on the underside of your forearm between your thumb and index finger. What is immediately apparent is the formation of ridges the are all equally spaced by some distance $\lambda$ and all seem to have approximately the same amplitude $A$ (Fig.~\ref{fig-sect1}d). So, a natural question is: what sets the spacing of these wrinkles? The physics at play here don't include gravity or a fluid, and balancing bending and stretching alone is not enough. This is a buckling problem, but one where a stiff film (skin) is resting on a softer substrate which is resisting deformation --  {\em i.e.} we can consider the mechanics of a beam on an elastic foundation~\cite{Dillard2018}. Consider the outer portion of your skin to be a stiff plate resting on a substrate that behaves like a Winkler foundation~\cite{Dillard2018}. We can write the strain energy of the foundation as
\begin{equation}
\label{Uf}
\mathcal{U}_f \sim K \int \delta^2 \ \dd \omega,
\end{equation}
where $K$ is the stiffness of the foundation, and $\delta$ is its deflection. A wrinkle will adopt a vertical deflection $\delta$ relative to some unknown characteristic spacing $\lambda$. Therefore, the curvature of these wrinkles which costs bending energy will scale as $\kappa \sim \delta/\lambda^2$. Balancing $\mathcal{U}_b$ with equation~\ref{Uf}, we expect that the wrinkles will have a wavelength of~\cite{Cerda2003}
\begin{equation}
\label{wrinkleLam}
\lambda \sim \left(\frac{B}{K}\right)^{1/4}.
\end{equation}
The physical interpretation of this scaling is that the stiff skin would like to buckle like an Euler column, with the lowest eigenmode. However, the substrate below would have to be stretched unevenly -- more at the center of the skin than near its edges. Therefore, the skin and the substrate reach a compromise set by equation~\ref{wrinkleLam}. The substrate does not need to be solid, indeed a fluid will impart an effective elastic stiffness of $K \sim \rho g$, where $\rho$ is the density of water~\cite{Huang2007, Holmes2010}.  For instance, if you consider placing a droplet of water on a thin film floating on a bath of water, wrinkles will form on the surface due to surface tension~\cite{Huang2007, Vella2010b, Huang2010, Davidovitch2011, Schroll2013}. This simple scaling breaks down when the loading is increased~\cite{Holmes2010}, as the wrinkles can no longer be treated as a small perturbation from the initially flat state, {\em i.e.} as the deformation transitions from the post--buckling {\em near--threshold} regime to the {\em far--from--threshold} regime~\cite{Vella2018}\footnote{This topic is discussed a bit more in the Supplemental Information}.


\subsection{Geometric Rigidity}
Before we learn about how to control and direct the deformations of slender structures, it is useful to understand where these structures derive their rigidity from. It has long been known that geometry alone can provide functionality, such as enhanced structural integrity -- arches have been used in architecture to bear loads for over four thousand years, and these structures exhibit their rigidity due to their intrinsic curvature. Shell structures, such as domes, are no different.  Research by Vella {\em et al.}~\cite{vella2011,vella2012a, vella2012b} and Reis {\em et al.}~\cite{lazarus2012geometry} has shed light on the intimate connections between a shell's geometry and it's mechanical behavior, demonstrating that tuning a shell's shape, rather than the materials, provides a straightforward way to enhance its ability to sustain a load. Consider a positively curved (\eg spherical, ellipsoidal), pressurized shell loaded at its apex by a point force. The rigidity of the shell depends on both the depth of indentation relative to shell thickness, and the degree of shell pressurization. In the limit of weakly pressurized shells, the classical results from Reissner are recovered~\cite{vella2011, vella2012a, lazarus2012geometry}, which find that the stiffness is dependent on the shell's Gaussian curvature $\mathcal{K}$ -- the shell's rigidity is correlated to the in--plane stretching of the shell~\cite{vella2012a}, an energetically costly deformation. Consider an egg shell, which is effectively ellipsoidal. If we take the Gaussian curvature at any point on the shell to be the product of the two principle curvatures, then $\mathcal{K}$ will be largest at the north pole of the shell, and lowest at the equator -- therefore the enhanced stiffness observed with compressing a chicken egg at its poles compared to its equator is a consequence of geometry--induced rigidity~\cite{lazarus2012geometry} (Fig.~\ref{fig-sect1}e). In the limit of high pressure, the mean curvature $\mathcal{H}$, not the Gaussian curvature, governs the shell stiffness~\cite{vella2012b}. The absence of a dependence on $\mathcal{K}$ in the large deflections of highly pressurized shells implies that the internal pressure negates the effect of geometric rigidity.

Understanding how a shell's stiffness depends on both the degree of internal pressure and the extent of deformation may be an important tool for biomechanical characterization. An important question for quantifying the morphogenesis of cellular and multicellular structures is how to deconvolute measurements of the cell wall mechanics from measurements of the pressure caused by osmotic fluid flow through the cell wall~\cite{Milani2013}, \ie turgor pressure. Vella {\em et al.}~\cite{vella2012a} demonstrated that their results on the indentation of pressurized elastic shells could characterize the turgor pressure within yeast cells, {\em viz. Saccharomyces cerevisiae}. Using the stiffness from the Reissner limit, a turgor pressure within the yeast cells was estimated that was consistent with those measured using other experimental techniques~\cite{Minc2009}. Preliminary work has begun on extending this analysis to the measurement of the elastic properties of tomato fruit cells~\cite{Zdunek2013} and plant tissues~\cite{Routier2013, Robinson2013}, and further development of this model to include different loading types may provide insight into the large deformations observed within artificial biological microcapsules, for example see~\cite{Neubauer2014} and the references therein.

\section{Elastic Instability Phenomena}
\label{stability}

The scaling from equation~\ref{U} tells us that thin structures prefer to bend rather than stretch, and in doing so will often exhibit an elastic instability. The efficient design of thin and lightweight structures out of high--strength materials leads to a conundrum at the heart of structural engineering: an optimum design is by its very nature prone to instability. With ample evidence that structural instability plays an integral role in the morphogenesis of biological materials~\cite{Goriely2017} -- from fingerprint formation~\cite{Groenewold2001, Kucken2005}, to the folds in the cerebral cortex~\cite{Goriely2015, Budday2015, Tallinen2016}, to the tendril perversion in climbing plants~\cite{Goriely1998} -- it is perhaps more important than ever that scientists and engineers studying solid materials have a strong understanding of the mechanics of elastic stability. 

Stability requires us to consider the internal exchange of energy within a structure, with the total potential energy consisting of the internal strain energy and the potential from the external loads, \ie $\mathcal{V}=\mathcal{U}_s+\mathcal{U}_b+\mathcal{P}$. We recall that the first variation of the total potential energy must be equal to zero for a structure to be in equilibrium, $\delta \mathcal{V}=0$. This statement is equivalent to Newton's second law. Equilibrium does not ensure stability, however. It is possible to balance a ball resting on the apex of a steep hill, such that it is in equilibrium, but this equilibrium is {\em unstable} because any slight perturbation will cause the ball to role far away and not return. A ball resting at the bottom of a hill is of course {\em stable}, because any perturbation to it, for example rolling it slightly up the hill will cause it to return to its original position once the perturbation is removed. We can also speak of {\em neutral} equilibrium, which would refer to the ball resting on a flat surface -- any perturbation will not change the ball's potential energy. With this simple analogy, we may recognize that these hills represent a potential energy landscape, and that we need to investigate the convexity or concavity of this landscape at an equilibrium point to determine if this equilibrium is stable or unstable, respectively. This requires us to investigate the character of the second variation of the total potential energy -- a structure is stable if $\delta^2 \mathcal{V}>0$, unstable if $\delta^2 \mathcal{V}<0$, and neutral if $\delta^2 \mathcal{V}=0$~\cite{HolmesSI}. It is tempting to relate these ideas to finding the extrema and curvature of a function in ordinary calculus, however energy is not a function, but rather a {\em functional} --  a function of a function of a variable. Evaluating the character of a functional requires the tools from {\em variational calculus}, and perhaps the most enjoyable primer on the calculus of variations can be found in Feynman's lectures~\cite{Feynman1977}.

For an elastic structure under a conservative load the critical point at which an instability occurs will always correspond to either a {\em bifurcation} (``buckling'') or a {\em limit point} (``snapping''). Imperfections in the structure's shape, the eccentricity of the load, or within the material can cause bifurcation critical points to change in character to a limit point, suggesting that buckling is perhaps more the exception than the rule~\cite{Budiansky2013}. On the surface, then it seems surprising that buckling problems are abound in the literature, while limit point problems appear far less frequently. There are, in general, two reasons for this: (1.) the distinction between bifurcations and limit points is not merely semantic: one mathematical tool that allow us to study bifurcations -- namely, linear stability analysis -- is impossible to use to study a limit point instability~\cite{Thompson1973, Thompson1984}, making buckling problems far easier to analyze, and (2.) slender structures are most efficient if they carry their load in a primarily {\em membrane} state of stress, and such configurations will typically fail by buckling rather than snapping~\cite{Budiansky2013}. The primary mathematical difference between a bifurcation and a limit point instability is this: a limit point occurs when the equilibrium position becomes unstable, and the structure moves to the closest, stable point {\em on the same equilibrium path}, while a bifurcation occurs {\em when two equilibrium pathways intersect}, and an exchange of stability occurs as the structure follows the stable equilibrium path. Concrete examples of these two generic instability phenomena are given in the Supplementary Information. For the chemist or physicist, these definitions may bring to mind a connection between these instabilities and {\em phase transitions}~\cite{Kosterlitz1973, Mikulinsky1995, Savel2004}, wherein one could draw an analogy between a limit point instability and a first--order phase transitions, and a bifurcation with a second--order phase transition. This is perhaps more useful conceptually than practically, but several similarities are shared. 

\subsection{Snapping}

The snap--through instability presents an important mechanism for directed shape change in the design of shape--shifting structures.  A recent review on this topic by Hu {\em et al.} will hopefully provide a nice compliment to the following discussion~\cite{Hu2015}.  There have been demonstrations of actuating snap--through instabilities for just about every conceivable mechanical and non--mechanical stimulus, including temperature~\cite{Jakomin2010}, light~\cite{Shankar2013}, acoustic excitation~\cite{Ng1989, Murphy1996}, elastomer or gel swelling \cite{holmes07, xia2010, Abdullah2016, Pezzulla2018}, magnetic fields~\cite{Zhu2013, Seffen2016b}, fluid flow~\cite{Gomez2017b}, surface tension or elastocapillarity~\cite{Fargette2014}, and electrical current with materials that include from ceramic (piezoelectric)~\cite{Schultz2003, Giannopoulos2007a}, metallic (electrostatic)~\cite{Zhang2007, Krylov2011}, and rubber (dielectric elastomers)~\cite{Keplinger2011, Zhao2014, Bense2017, Shao2018}.  Laminated composites of epoxy and carbon fiber or fiber glass may exhibit bistability or multistability while thermally curing~\cite{Hufenbach2002, Lachenal2012, Zhang2013, Ryu2014}. Depending on the fiber orientation, the presence of the fiber embedded in the epoxy matrix may give these materials an orthotropic response, thus providing design criteria for the orientation of the stimulus chosen to induce snapping~\cite{Seffen2007}. Such structures may enable morphing components for wind turbines~\cite{Lachenal2013}, mechanisms for swimming or flying~\cite{Lienhard2011}, ventricular assist devices~\cite{goncalves2003}, and for lightweight, deployable structures~\cite{Seffen1999, Walker2007}. In a similar manner, the orientation of the residual stress or prestress resulting from fabrication can alter the geometric criteria for bistability~\cite{Kebadze2004, Chen2012a, Jiang2018}. In addition, rapid prototyping techniques, such as 3D printing and laser cutting, have made it much easier to generate bistable structures on a wide range of scales. Of note is the elastomer coating technique pioneered by Brun {\em et al.}, which has made the preparation of thin, spherical caps with nearly uniform thickness simple, fast, and affordable~\cite{Brun2016}. With all of these ways to induce snapping, it is no surprise that it has found utility in a wide range of engineering fields. 
 
Experimental results on the adhesion of 2D materials like graphene and MoS2 have indicated that there is a snap into adhesive contact between the film and a substrate containing roughness~\cite{Li2009, Scharfenberg2012}, which has enabled researchers to demonstrate that this instability may be a useful metrology tool to measure the material properties of 2D materials~\cite{Lindahl2012}. The snap--through instability has also been utilized to tune a material's properties in response to an applied load by altering the material's lattice structure to generate dramatic, dynamic increases in a material's stiffness~\cite{Jaglinski2007}. On a slightly larger scale, there is interest in the micro- and nanoelectromechanical (MEMS and NEMS, respectively) communities to use the rapid snap--through of arches and shells in electromechanical systems~\cite{hsu2003} for accelerometers~\cite{Hansen2007}, or as a means to rapidly change a surface's texture or optical properties~\cite{holmes07}. The precise placement of folds in thin sheets can generate a wide range of multistable structures, with the most fundamental being the {\em waterbomb}~\cite{Hanna2014bh, Bowen2015, Chen2016, Brunck2016}, which has generic bistability for any number of creases~\cite{Brunck2016}. In addition to traditional folding and cutting techniques, programming creases into a material through spatial variations in its thickness can enable bistability in folded shells -- cylinders, spheres, or saddles~\cite{Bende2015}. One function that has drawn particular attention is the design of bistable structures as a means to capture, trap, or harness elastic strain energy~\cite{Pellegrini2013, Shan2015, Raney2016}. This process may be amenable for use with soft elastomers~\cite{Shan2015}, which can undergo very large snap--through instabilities without exhibiting material failure~\cite{Keplinger2011, Li2011, Zhao2014, Shao2018}. If these soft materials are used as actuators, rather than energy harvesters, the instability can be used to trigger large, nonlinear changes in shape, pressure, and extension within soft, fluidic actuators~\cite{Overvelde2015}.

\subsection{Buckling}

The elastic buckling of a beam or plate provides a straightforward way to get large, reversible, out--of--plane deformations which can be utilized for generating advanced functionality. For instance, embedding a flexible plate within a microfluidic channel provides a means for mechanically actuated valving~\cite{Holmes2013, Tavakol2016}. Buckled plates have been utilized to fabricate semiconductor nanoribbons for stretchable electronics~\cite{Someya2005, Sun2006, Lu2006}.  In recent years~\cite{Kim2008b, Kim2008c, Khang2009}, researchers have utilized this principle to fabricate single--wall carbon nanotube arches~\cite{Khang2008}, single-crystalline silicon arches that were used as metal-oxide semiconductor field--effect transistors (MOSFETs)~\cite{Kim2008}, and buckled lead zirconate titanate (PZT) ribbons for ceramic piezoelectric energy harvesting devices~\cite{Qi2011}. Buckling bilayer plates can be utilized to generate shape--shifting structures~\cite{Pezzulla2016} that may be used as soft grippers~\cite{Abdullah2018}. Buckling of shells has provided an intriguing way to control global shape~\cite{Knoche2011, Nasto2013, Pezzulla2018} and local patterning~\cite{Stoop2015, Jimenez2016, Marthelot2017, Chung2018}, reduce aerodynamic drag~\cite{Terwagne2014, Guttag2017}, generate lock--and--key colloids that can selectively aggregate~\cite{Sacanna2011}, form liquid crystal shell actuators~\cite{Jampani2018}, and pave the way for buckling microswimmers~\cite{Djellouli2017}. 

Wrinkles are what appear when a thin structure supported by a soft substrate buckles~\cite{Allen1969, Cerda2002, Cerda2003, Efimenko2005, Huang2007, Li2012}. There are a variety of ways to fabricate wrinkled surfaces, but the underlying principle is simple: bond a stiff film containing residual compressive stress onto a compliant substrate. Some of the first experiments in this manner involved the deposition of thin metallic film onto a PDMS substrate by electron beam evaporation~\cite{Bowden1998, Huck2000}. This deposition locally heats the PDMS surface, expanding it equibiaxially. Upon cooling, this compressive stress causes the metallic film to buckle with wrinkles in a herringbone pattern that cover a large surface area. Wrinkles have long added functionality to structural materials, such as with their ability to damp the structural vibrations occurring on composite plates~\cite{Parfitt1962}. Similar to the buckled ribbons used for flexible electronics, these wrinkled plates enabled the fabrication of stretchable gold conductors~\cite{Lacour2003}, flexible-electronic devices using wrinkled ribbons~\cite{Choi2007, Khang2009, Wang2011} and wrinkled single-walled carbon nanotubes~\cite{Yu2009}. Sinusoidal wrinkles have been used to alter and tune friction~\cite{Rand2009,Suzuki2014}, to fabricate tunable phase gratings~\cite{Harrison2004}, and to improve the metrology of thin films via the strain--induced elastic buckling instability for mechanical measurement (SIEBIMM) technique~\cite{Stafford2004}.  Additionally, there is significant evidence within biological systems to suggest that patterned surfaces enhance adhesion~\cite{Autumn2002, Chan2007, Davis2011, Davis2012}. Control of the pattern topography is a crucial component for utilizing these structured surfaces (Fig.~\ref{fig-overview}o). Linear stability of the cylindrical pattern reveals the emergence of hexagonal, triangular, and square modes, and the commonly observed herringbone pattern~\cite{Chen2004b, Audoly2008a, Cai2011}. Finally, as the amount of overstress, or confinement, of the compressed plate increases, the bending energy along the plate goes from being broadly distributed among wrinkles to being localized within sharper features~\cite{Pocivavsek2008, Holmes2010, HolmesSI}, as the deformation transitions from the post--buckling {\em near--threshold} regime to the {\em far--from--threshold} regime~\cite{Vella2010b, Davidovitch2011, Vella2018}.

\section{Programmable Matter}
\label{programmable}

\subsection{Mechanical Metamaterials}

\begin{figure} 
\begin{center} 
\includegraphics[width=1\columnwidth]{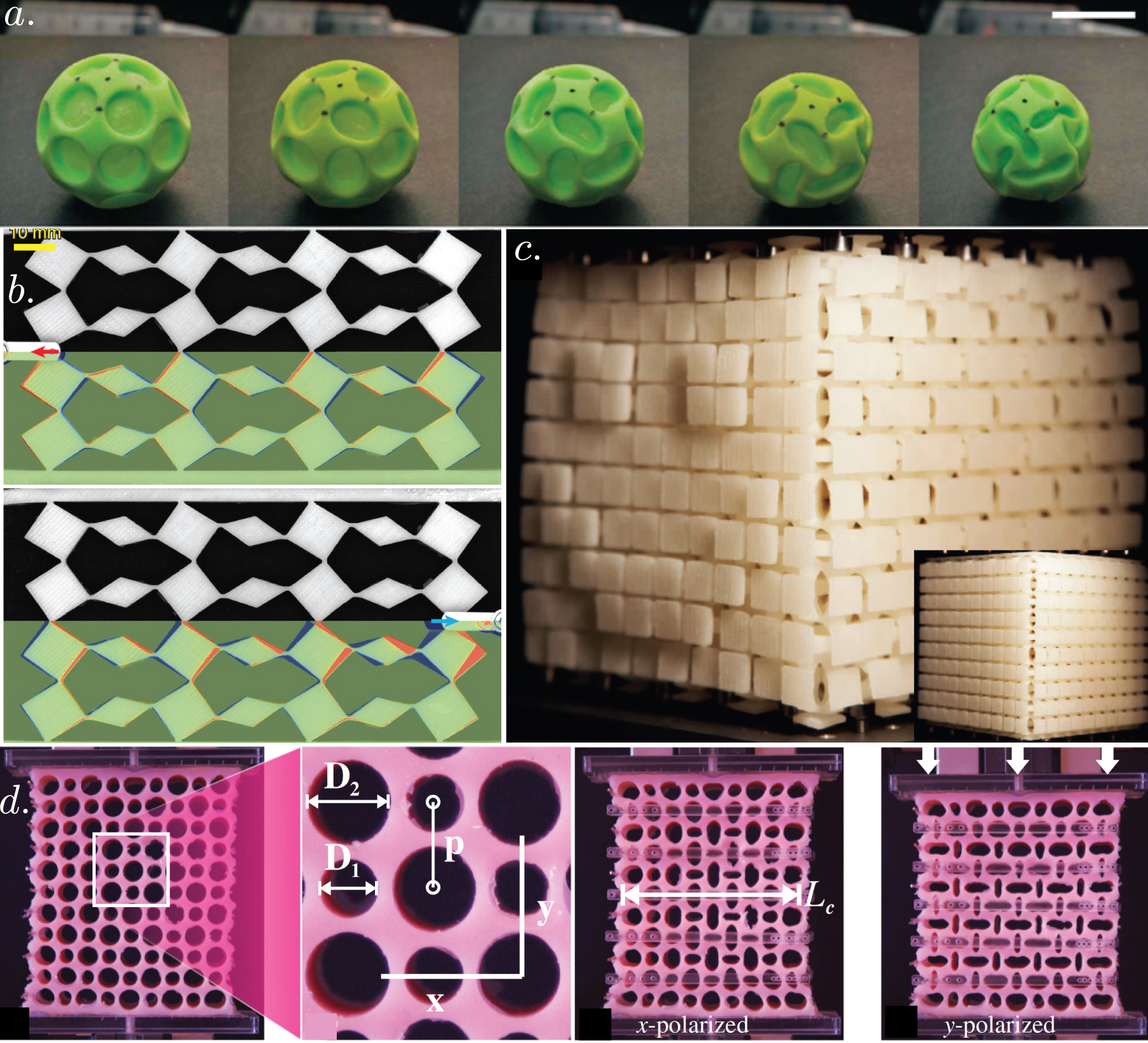} 
\end{center}
\vspace{-7mm} 
\caption{$a.$ Spherical encapsulation with a ``buckliball'', adapted from~\cite{Shim2012}, $b.$ static non--reciprocity in mechanical metamaterials~\cite{Coulais2017}, $c.$ programmatic shape change with a pixelated cube, adapted from~\cite{Coulais2016}, and $d.$ programmatic structural polarity through external constraints, adapted from~\cite{Florijn2014}.  \label{fig-mechmeta}} 
\end{figure}

Metamaterials are rationally designed structures composed of building blocks to enable functionality that cannot be found in natural materials~\cite{Kadic2013}.  While this term has been traditionally associated with electromagnetism and optics, the field has recently broadened to include elastostatic and elastodynamics metamaterials, collectively called {\em mechanical metamaterials}. There is an excellent review that will cover the details of this topic far better than I can will do here, so I encourage the curious reader to seek out Bertoldi \etal~\cite{Bertoldi2017}. In general, the underlying concept of a mechanical metamaterial is to use hierarchical structures, such as microscale or mesoscale geometric features, to alter the macroscopic deformation of an object. Foam has natural microscale features, and its atypical mechanical response exhibits a negative Poisson ratio~\cite{Lakes1987}. This auxetic behavior~\cite{Grima2000} has been explored in a variety of contexts, many of which utilize the buckling of micro/mesoscale beams within the macroscale structure~\cite{Mullin2007, Bertoldi2010}. With the inclusion of additional constraints, this auxetic behavior can be finely tuned to provide a programmatic structural deformation with a tailored force--displacement response~\cite{Florijn2014, Florijn2016} (Fig.~\ref{fig-mechmeta}d). Buckling of microstructural features to generate auxetic behavior can be extended from structured plates to structured shells to potentially encapsulate material~\cite{Shim2012} (Fig.~\ref{fig-mechmeta}a). Instability is at the heart of mechanical metamaterials that exhibit negative compressibility, in which they contract while being loaded in tension, potentially enabling force amplification~\cite{Nicolaou2012}. Buckling of subscale features becomes truly ``meta'' when one considers the Euler buckling of metabeams~\cite{Coulais2015}, wherein the microstructure has a significant affect on the postbuckling response. More generally, precise control of {\em pixelated} structures provides a way to programmatically dictate a structure's deformation across multiple length scales~\cite{Coulais2018} (Fig.~\ref{fig-mechmeta}c). For example, the combinatorial design of specific building blocks can enable a cube to produce a smiling face in response to uniaxial compression~\cite{Coulais2016}. Mechanical metamaterials are now veering closer to classical metamaterials by examining the role of topology in addition to geometry to transform shape and mechanical properties~\cite{Rocklin2017}, with notable examples including structures to break the static reciprocity formalized by the Maxwell--Betti reciprocity theorem~\cite{Coulais2017} (Fig.~\ref{fig-mechmeta}b).

\subsection{Origami \& Kirigami}

Recall that Gauss's famous {\em Theorema Egregium} states that you cannot change the intrinsic curvature of a surface without stretching it, and we know from equation~\ref{U} that it is far easier to bend a thin object than stretch one. So, if you want to stretch a thin object or wrap a sheet of paper around a soccer ball, what is one to do? One answer is to design regions that enable stretching, such as with the inclusion of folds and cuts. A review by Witten~\cite{Witten2007}, which in my opinion did much to catalyze this entire discipline, describes at great lengths the way stress within thin sheets will localize into sharp creases and folds, essentially acting as sacrificial regions in which a material begrudgingly stretches. Understanding the mechanical nature of these folds guides their strategic incorporation into engineering systems. If these folds are precisely placed and sequentially actuated, they represent a means to develop advanced engineering structures. Taking inspiration from the Japanese art of {\em origami}, sequential folding has long been utilized in structured systems~\cite{Engel1981}. Deployable structures, utilized in space technologies, typically require damage--free actuation, reliable deformation, and autonomous or automated conformational change, and have been utilized within small satellite deployable structures, deploy booms, and array panels~\cite{Pellegrino2001, Walker2007}. Recent research has examined the mechanics of these foldable structures, with a focus on origami--inspired design~\cite{Cipra2001, Klett2011}, and thus it may be useful to the reader to seek out the review article by Peraza--Hernandez \etal on this specific topic~\cite{Peraza2014}. The fundamental mechanics of a fold~\cite{Lechenault2014}, or more broadly a conical defect or singularity within a thin plate~\cite{Farmer2005, Muller2008, Guven2013, Seffen2016a} are at the heart of understanding and designing origami mechanisms. While the geometric properties of a fold have been studied in great detail, the role of a material's properties in these systems has been largely overlooked to date. A notable exception is the work by Croll \etal on the role of adhesion in crumpled structures~\cite{Croll2018}. Unit cells of a repeating fold pattern provide origami building blocks~\cite{Waitukaitis2016}, and enable the programmatic design of structural deformations~\cite{Silverberg2014, Filipov2015, Overvelde2017}, 3D shape--shifting~\cite{Overvelde2016}, and folding--induced curvatures~\cite{Dudte2016}. Such systems may exhibit multistability, as noted briefly in section on snapping~\cite{Waitukaitis2015, Silverberg2015}. 

The Miura-ori folding pattern is a well--known example of functional origami -- a parallelogram of folds defined by two fold angles and two lengths, which enables the compact folding of a flat plate~\cite{Miura1980, Wei2013, Schenk2013}. This functional array of pleated folds has been utilized for collapsable maps, deployable satellite arrays~\cite{Zirbel2013}, and to increase the areal energy density of folded lithium--ion batteries~\cite{Cheng2013}.  Mathematicians and mechanicians alike have borrowed concepts from origami to create biomedical stents~\cite{You2003}, nanostructured foldable electrodes~\cite{In2006}, ultrathin high--resolution, folded optical lenses~\cite{Tremblay2007}, photovoltaics~\cite{Myers2010}, materials with tunable coefficients of thermal expansion~\cite{Boatti2017}, and for folding rigid, thin--walled structures around hinges~\cite{Wu2011}. The techniques for generating folds include actuation by shape--memory alloys~\cite{Hawkes2010}, light~\cite{Ryu2012, Liu2012}, microfluidic flow~\cite{Liu2011}, and direct--printed wet--origami~\cite{Ahn2010}.

A natural extension of using folds inspired by {\em origami} to enable elaborate shape changes is to use cuts inspired by {\em kirigami}. Here, the fundamental behavior is governed by the nonlinear mechanics of thin frames~\cite{Moshe2018}. With kirigami, the roles of topology~\cite{Chen2016top} and geometry are once again at the heart of design rules that enable targeted shape changes~\cite{Castle2014, Sussman2015, TangYin2017}. Mechanics enters the picture by enabling these flat sheets to buckle into 3D structures when stretched or compressed beyond a critical point~\cite{Ahmadkatia2017, Dias2017}. While lattice cuts have primarily been studied to date, with the work of Dias \etal being a notable exception~\cite{Dias2017}, additional functionality was demonstrated with multiscale cuts, \ie {\em kiri--kirigami}~\cite{YTang2017}, randomly oriented~\cite{Grima2016} and fractal cuts~\cite{Cho2014, YangS2016}, and patterns that naturally interlock~\cite{Wu2016}. One of the primary demonstrations of functional shape change via kirigami is the extreme stretchability of thin sheets~\cite{Shyu2015, Zhang2015, YTang2015}, which even translates to 2D materials, such as graphene~\cite{Blees2015}. Dias \etal showed that to leading order, the out--of--plane deformation of a single cut is independent of the thickness of the sheet, providing mechanical insight into why this shape changing mechanism can scale down to the thinnest materials~\cite{Dias2017}. Kirigami--inspired cut patterns have enabled the fabrication of photonic metamaterials~\cite{Ou2011}, metamaterials with tunable material properties~\cite{Hwang2018a, Yang2018}, smart adhesives~\cite{Hwang2018b}, solar tracking devices~\cite{AaronL2015}, stretchable lithium ion batteries~\cite{Song2015}, optical beam steering~\cite{Wang2017}, shape--shifting structures~\cite{Robin2016}, and ultralightweight linear actuators~\cite{Dias2017}. 

\subsection{Shape--Shifting Structures}


\begin{figure} 
\begin{center} 
\includegraphics[width=1\columnwidth]{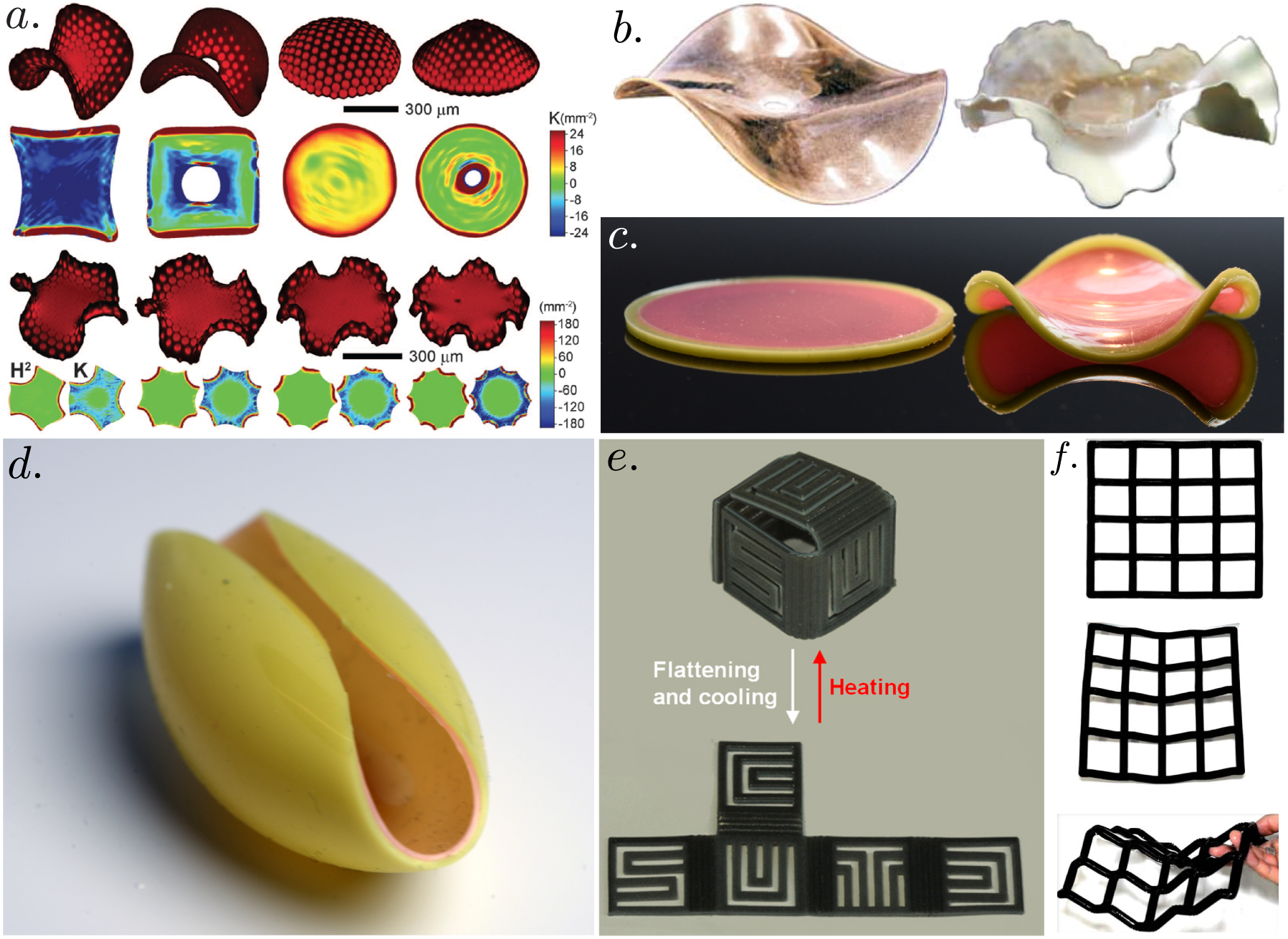} 
\end{center}
\vspace{-7mm} 
\caption{$a.$ Morphing with halftone lithography adapted from~\cite{Kim2012a}, $b.$ non--Euclidean plates adapted from~\cite{Klein2007b}, $c.$ a monkey saddle, adapted from~\cite{Tibbits2017}, $d.$ pollen grain isometry of a spherical cap, related to the work in~\cite{Pezzulla2018}, $e.$ 4D printing using shape memory polymer joints, adapted from~\cite{Ge2014}, and $f.$ active printed meshes, adapted from~\cite{Raviv2014}.  \label{fig-swell}} 
\end{figure}

The control of geometry and elasticity to create adaptive, morphing structures paves the way for an era of {\em designer materials}~\cite{Reis2015}. We have reviewed the energetic limitations on bending and stretching structures, and we have seen how carefully chosen stretching regions -- through, for example, origami and kirigami -- enable a much wider range of shape changes. An alternative approach for shape--shifting structures is through the programmatic design of the volumetric strain within a material. This approach draws the closest analogy to synthetic structural growth. Swelling presents a simple and effective technique to spatially tune volumetric strain, while falling short of growth by not permanently adding mass to the reference elastomer, or reference configuration. Swelling is the infiltration of an elastic network with a favorable solvent, such that the characteristic size of the polymer chains, \ie their radius of gyration, dramatically increases. The balance of fluid movement and elastic response dictates the time scale over which swelling--induced phenomena will occur, and the ability for the fluid to swell the polymer network dictates the magnitude of stress imparted. Fluid-flow within natural structures plays a crucial role in the morphology of growing tissues~\cite{Moulia2000, Dervaux2009, Dervaux2011}, the opening of seed pods~\cite{Armon2011}, the shrinkage of mud~\cite{Shorlin2000} and moss~\cite{Arnold1899}, and the curling of cartilage~\cite{Setton1998}, leaves~\cite{King1996, Marder2003}, and pine cones~\cite{Reyssat2009}. Porous thin films, such as fuel cell membranes~\cite{Zhou2009}, are highly susceptible to swelling-induced delamination and buckling, which cripple their functionality.

Swelling--induced deformations provide a means for shaping two-dimensional sheets into three-dimensional structures~\cite{Klein2007b, Holmes2011, Kim2012a}, with features spanning multiple length scales~\cite{Dervaux2012, Pandey2013}. Differential swelling -- where portions of the material locally swell more than others -- have been used to create helical ribbons~\cite{Wu2013}, cylinders~\cite{Pezzulla2016}, saddles~\cite{Pezzulla2015}, pinched spheres~\cite{Pezzulla2018}, and wavy strips and discs~\cite{Mora2006, DuPont2010, Barros2012}. The shape selection process is nontrivial, and many efforts have focused on predicting what shape will emerge from  {\em programming} the volumetric strain, often tailored using the metric tensor of the middle surface of a thin plate or shell~\cite{Efrati2009a, Santangelo2009, Sharon2010, Gemmer2011, Pezzulla2017, Pezzulla2018}. The alternative, inverse problem -- predicting what metric to program to achieve a specific 3D shape -- may be even more desirable~\cite{Dias2011}. In addition to the contributions from the sheet's elastic energy, understanding the dynamic morphing of a swelling structure, such as the curling of paper~\cite{Douezan2011, Reyssat2011}, gels~\cite{Kim2012b}, and rubber~\cite{Holmes2011, Nah2011}, poses additional challenges~\cite{Lucantonio2012, Lucantonio2013}. An intricate photolithography technique, {\em e.g. halftone lithography}, has been developed to scale this dynamic process down to create responsive, morphing structures on the micron scale~\cite{Kim2012a}. The mechanics behind this structural morphing combines dynamic aspects of swelling, structure, stability, and material properties, thereby creating a rich environment for experimental and theoretical insights. It is likely that building on the concept of {\em programmable matter}, will inspire novel rapid--prototyping technologies, such as 3D ``inkjet'' printing that utilizes small amounts of polymerizable solvents to create complex structures. Such ideas are already emerging in a technique known as {\em 4D printing}, where the fourth dimension represents the timescale corresponding to swelling~\cite{Tibbits2014, Ge2014, Raviv2014, Khoo2015, Bakarich2015, Gladman2016, Huang2017, Ding2017}.

\section{Summary \& Outlook}

Shape changing materials require working with and around the constraints of elasticity, and utilizing nonlinearity to generate functionality. Most of what I have covered in this review does exactly that -- overcoming the difficulty of stretching thin sheets by folding or cutting them, using the buckling and snapping of beams and plates and shells to generate metamaterial behavior, and playing with swelling to tune spatial distributions of strain. So, where do we go from here? Certainly, important research will continue in the areas we have covered at length, though new opportunities are beginning to emerge in a few specific areas. I think one of the most significant areas of research going forward will focus on woven fabrics and knits for tailoring shape and properties~\cite{Poincloux2018, Dimitriyev2018}, and for example, such ideas are already enabling the form finding of grid shells~\cite{Baek2018}. Another interesting area may lie at the interface between elastic structures and granular or colloidal materials, so--called {\em elastogranular} interactions. Research in this regard is beginning to lay out the fundamental behavior of bending and packing of elastic rods within grains~\cite{Mojdehi2016, Schunter2018, Algarra2018}, which is clearly relevant in the form and function of growing plant roots~\cite{Kolb2012, Wendell2012, Kolb2017}, and is beginning to show promise for programmable, reversible architecture~\cite{Aejmelaeus2016, Dierichs2016, Fauconneau2016}. Finally, in my opinion the biggest limitation in achieving the next generation of shape--shifting structures is the absence of simple to fabricate and robust materials that are highly responsive to stimuli -- \ie we need help from chemists and materials scientists. Dielectric elastomers offer a lot of promise, yet require extreme voltages and fail often. Swelling elastomers and gels is slow, requires a bath of fluid, and usually involves brittle materials that fail when they either deform too much or dry out. We need addressable and programmable materials to take full advantage of the recent advances in mechanical metamaterials. Too often the research covered in this review utilizes ordinary office paper, dental polymers, and traditional acrylic plastics and urethane rubbers. This approach is fine, and should even be applauded, when the purpose is to show how an appropriate understanding of mechanics, geometry, and topology can make profound predictions with run--of--the--mill materials, but additional materials science advances are necessary to fully realize the potential opportunities for technological insertion of shape--shifting materials. Among the leaders of this effort to connect materials and mechanics is the work of Hayward {\em et al.}, see for example~\cite{Hauser2015, Hauser2016, Jeon2017}. Another recent interesting effort to tune the material response of traditional thermal bimorphs highlights the ability to use encapsulated phase change materials to spontaneously induce thermal bending at critical temperatures~\cite{Blonder2017}. There is much work to be done to better blend the insights from mechanics with the advances in materials.

\section{Acknowledgements}

I gratefully acknowledge the financial support from NSF CMMI -- CAREER through Mechanics of Materials and Structures (\#1454153). I am grateful to L. Stein--Montalvo and X. Jiang for photos, and I would like to thank M.A. Dias for many discussions on this topic over the years. I also thank the organizers of the Kavli Institute for Theoretical Physics (NSF PHY-1748958) program on Geometry, Elasticity, Fluctuations, and Order in 2D Soft Matter for providing me an opportunity to think through some of the ideas detailed within this article. I apologize if I did not cite your work in this review, it was an accident and an oversight on my part, don't @ me.








\newpage

\noindent{\Large \textsc{Supplementary Information:}}

\noindent{\large Elasticity and Stability of Shape Changing Structures}

\setcounter{section}{0}
\setcounter{equation}{0}
\setcounter{figure}{0}
\renewcommand{\thesection}{\Alph{section}}

\section{A bit more on stretching \& bending}

Shell theories have a common structure: a stretching energy $\mathcal{U}_s$, which accounts for in--plane deformations, that is linear in the shell thickness, $h$, and a bending energy $\mathcal{U}_b$, which accounts for the curvature change of the deformed shell, and is dependent on $h^3$~\cite{Koiter1960, Koiter1966, Koiter1967, Koiter1973}. The dimensionless strain energy of the Koiter shell equations are written in covariant form as~\cite{Niordson1985}
{\setlength{\mathindent}{0cm}
\begin{subequations}
\begin{align}
\label{Us}
\mathcal{U}_s&= \frac{Y}{2} \int \left[(1-\nu)\varepsilon^{\alpha\beta}\varepsilon_{\alpha\beta}+\nu(\varepsilon_\alpha^\alpha)^2\right] \dd\omega \sim Y \int  \varepsilon^2 \dd \omega, \\
\label{Ub}
\mathcal{U}_b&=\frac{B}{2} \int  \left[(1-\nu)\kappa^{\alpha\beta}\kappa_{\alpha\beta}+\nu(\kappa_\alpha^\alpha)^2\right] \dd\omega \sim\! B  \int \kappa^2 \dd \omega,
\end{align}
\end{subequations}}where $Y=Eh/(1-\nu^2)$ is the stretching rigidity, $B=Eh^3/[12(1-\nu^2)]$ is the bending rigidity, $E$ is Young's elastic modulus, $\nu$ is Poisson's ratio, $\dd \omega$ is the area element, and $\vett{\varepsilon}$ and $\vett{\kappa}$ are the tensors that describe the stretching and bending of the middle surface of the shell, respectively. The Greek indices $\alpha \in [1,2]$, and raised indices refer to contravariant components of a tensor, while lowered indices refer to covariant components. The intricacies of this notation are not important upon a first introduction to this topic, but for the curious reader I would recommend the instructive text by Niordson~\cite{Niordson1985}. The energies each contain the square of two invariants, since the shell is two--dimensional, for example with $\varepsilon_\alpha^\alpha$ being a measure of how the area changes and $\varepsilon^{\alpha\beta}\varepsilon_{\alpha\beta}$ being a measure of how the shell is distorted.

\section{A bit more on elastic stability}

Unfortunately, most student's first and only exposure to a problem of elastic stability is a rather deceptive one -- the Euler buckling of a column. In studying the Euler buckling of a column, it is customary to make a one particularly significant assumption -- that the column is {\em incompressible}. This assumption is advantageous because it is both reasonable, as most columns do not shorten significantly before buckling, and it simplifies the calculation. However, it is quite misleading, and I will let one of the forefathers of stability theory, W.T. Koiter explain why (emphasis mine): \vspace{-2mm}
\begin{myindentpar}{6mm}{6mm}
{\em ``Here it appears that a negative second variation of the potential energy of the external loads is the cause of a loss of stability, but this state of affairs is due to our simplifying assumption of an incompressible rod, and we may not generalize this experience to other cases, as still happens only too frequently. In fact, in the case of elastic structures under dead loads the potential energy of the external loads does not enter at all into the stability condition. {\bf Loss of stability of elastic structures is always due to an internal exchange of energy.}''}~\cite{Budiansky2013}
\end{myindentpar}
What that means physically is that it is the energy, and more specifically its second variation, in the {\em deformed} or {\em fundamental} state that needs to be evaluated for stability, and not the potential acting on the structure in its {\em original} or {\em reference} state. For a more fundamental and at times conceptual discussion of elastic stability, I recommend the two textbooks by Thompson and Hunt~\cite{Thompson1973, Thompson1984}.

\section{The simplest snapping structure}

\begin{figure} 
\begin{center} 
\includegraphics[width=1\columnwidth]{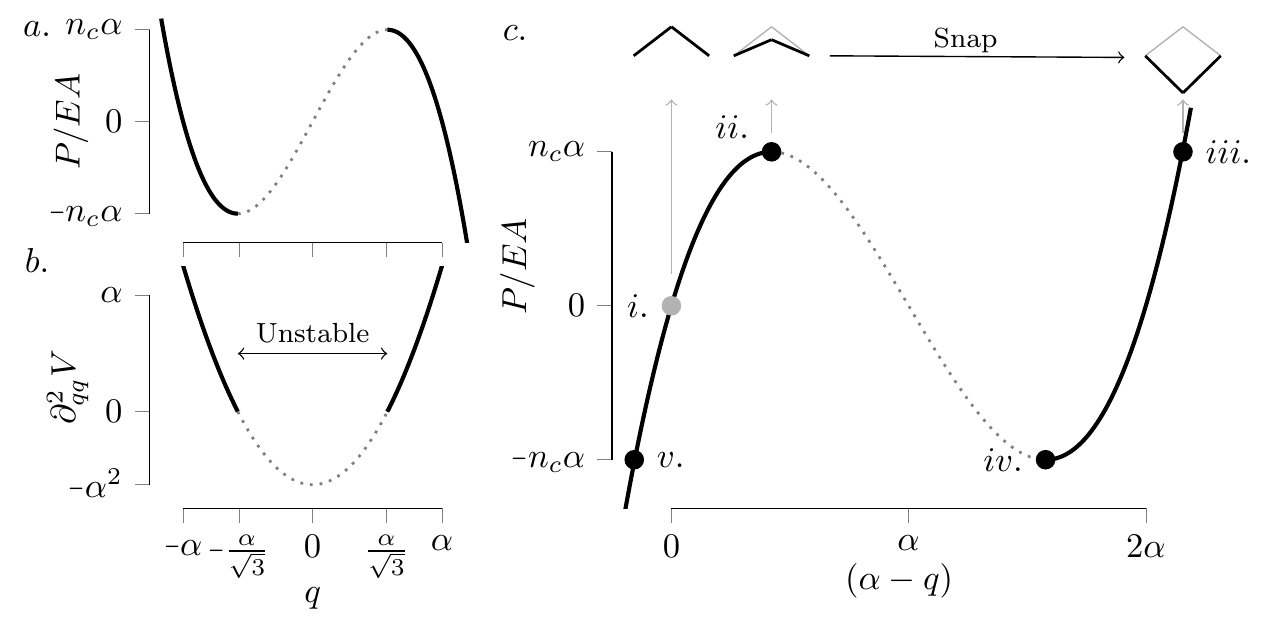} 
\end{center}
\vspace{-7mm} 
\caption{$a.$ Normalized force $P/EA$ plotted as a function of angle $q$ showing extrema at $n_c\alpha$, where $n_c=-\frac{2\pi^3}{75\sqrt{3}}$.  $b.$ The second variation of the total potential energy is plotted as a function of angle $q$. The equilibrium configuration is unstable when $\partial^2_{qq}\leq0$. $c.$ Normalized force is plotted as a function of rotation, showing the loading path from $i.$ the initial configuration to $ii.$ the limit point at $n_c\alpha$, at which point stability is lost and the structure snaps to $iii.$ Upon unloading from $iii.$ to the limit point at $iv.$ the structure snaps to $v.$ Removal of the load restores the initial configuration $i.$ \label{fig-snap}} 
\end{figure}

The simplest structure that exhibits snapping is the bistable truss first analyzed by von Mises, in which two elastic bars of elastic modulus $E$, cross sectional area $A$, and an axial stiffness $EA/(L/\cos{\alpha})$ are pinned together with an initial angle $\alpha$, and pinned at supports separated by a distance $2L$. A force $P$ applied to the apex of the truss will change the angle from $\alpha$ to $q$, and induce a compressive strain in the bars of $\varepsilon=\cos{\alpha}/\cos{q}-1$, which causes a vertical displacement $w=L(\tan{\alpha}-\tan{q})$. In this single degree of freedom example, the total potential energy is simply the strain energy in the spring minus the potential energy of the load, 
\begin{equation}
\label{eq-VonMisesV}
V(q)=\frac{EAL}{\cos \alpha}\left(\frac{\cos \alpha}{\cos q}-1\right)^2-PL(\tan \alpha-\tan q).
\end{equation}
Equilibrium is found by setting the first variation of the potential energy to zero,
\begin{equation}
\label{vonM-dV}
\frac{\partial V}{\partial q}=\frac{L}{\cos^2 q}\left[2EA(\cos \alpha \tan q - \sin q)+P \right] =0,
\end{equation}
from which we can find the equilibrium path
\begin{equation}
\label{vonM-P}
\frac{P}{EA}=2(\sin q -\cos \alpha \tan q),
\end{equation}
which is plotted in figure~\ref{fig-snap}a. Here we note an important distinction between snapping, which is a limit point instability, and buckling, which is a bifurcation -- there is only on equilibrium path, meaning the truss will deflection under any infinitesimal load $P$.  A system that exhibits a limit point instability has only one equilibrium path, and loss of stability will cause a discontinuous jump in a given parameter, {\em i.c.} a finite change in $q$ in response to an infinitesimal change in $P$. Stability of this truss will be lost when the second variation of its potential energy, $\delta^2V=\frac{1}{2!}\frac{\partial^2 V}{\partial q^2}\delta q^2$, ceases to be positive definite.  We will evaluate the stability of the equilibrium state at a constant load by inserting the load from equation~\ref{vonM-P} into equation~\ref{vonM-dV}, and calculating the second variation of $V$ as
\begin{equation}
\label{vonM-d2V}
\frac{\partial^2 V}{\partial q^2} =2 \frac{EAL}{\cos^4 q}\left(\cos \alpha -\cos^3 q\right),
\end{equation}
which is plotted in which is plotted in figure~\ref{fig-snap}b. It is clear from the graph that a region of this second variation is below zero, specifically when $-\alpha/\sqrt{3} \leq q \leq \alpha/\sqrt{3}$. This region of the equilibrium path given by equation~\ref{vonM-P} is unstable. Since in this example we have considered a load--controlled experiment, meaning the magnitude of the low is being increased, the truss has no choice but to jump (horizontally) from point $ii.$ to point $iii.$ on Fig.~\ref{fig-snap}c., since this is the closest, stable part of the equilibrium curve at the same fixed value of $P/EA$.

In general, finding the critical point of a snap--through instability is a challenge, because when you linearize the equations you lose all information about the instability. This can be immediately seen with our simple example of the von Mises truss from equation~\ref{eq-VonMisesV} -- a Taylor series expansion of $q$ for any $\alpha$ leaves you with a linear equation, and thus the critical point is lost -- its second variation will never be negative, so one would erroneously think the system is always stable. Numerically, there are multiple approaches for finding, following, and continuing through instability points~\cite{Riks1979}, including utilizing arc--length methods or dynamic simulations.

\section{The simplest buckling structure}

\begin{figure} 
\begin{center} 
\includegraphics[width=1\columnwidth]{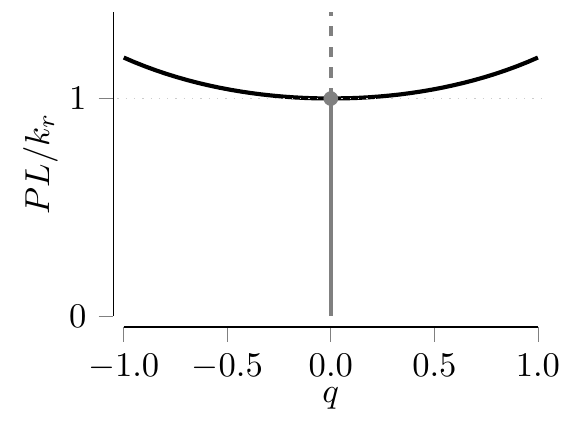} 
\end{center}
\vspace{-7mm} 
\caption{A plot of the two equilibrium curves of a buckling rigid bar, using the equilibrium paths given by equation~\ref{eq-Pq}. When $PL/k_r \leq 1$, the bar remains vertical, and it buckles either to the left ($q<0$) or right ($q>0$) when $PL/k_r > 1$. At the critical point, which is stable~\cite{Thompson1973, Thompson1984}, there is an exchange of stability between the gray equilibrium curve and the black curve as the structure buckles.  \label{fig-buck}} 
\end{figure}

To see the buckling and postbuckling behavior of a structure, it is useful to consider a nonlinear, finite--deflection theory -- something that is straightforward to examine in a discrete, one degree of freedom system. Consider a rigid bar of length $L$ that is held vertical by a rotational spring of stiffness $k_r$, and loaded at the free end of the bar by a a load $P$. If it is perfectly vertical, there is no initial inclination angle to the bar\footnote{To consider the role of imperfections, or to analyze the stability of a whole array of structures and materials, I recommend pouring over the invaluable tome by Ba{\v{z}}ant and Cedolin~\cite{bazant}, especially chapter 4.}, {\em i.e.} $\alpha=0$. We can again use $q$ to measure the the inclination angle of the bar after loading, so that the total potential energy may be written as
\begin{equation}
\label{eq-bar}
V(q)=\frac{1}{2}k_rq^2-PL(1-\cos q),
\end{equation}
where the first term on the right hand side is the strain energy stored in the rotational spring, and the second term is the work of the load. Its first variation yields
\begin{equation}
\frac{\partial V}{\partial q} = k_r q - PL \sin q,
\end{equation}
and since equilibrium requires the first variation to be stationary, {\em i.e.} $\partial V=0$, we find the equilibrium relation between force and angle to be
\begin{equation}
\label{eq-Pq}
P(q)=\frac{k_r}{L}\left(\frac{q}{\sin q}\right).
\end{equation}
Upon inspection of equation~\ref{eq-Pq}, we see that there are two solutions that satisfy this equation: $q=0$ or $q \neq0$ (Fig.~\ref{fig-buck}). The bar can stay perfectly vertical, {\em i.e.} $q=0 \ \forall \ P$, but that solution is not always stable. We can check the stability of our system by evaluating the second variation of $V$, and requiring it to be positive definite. The second variation is 
\begin{equation}
\label{eq-d2V}
\frac{\partial^2 V}{\partial q^2} = k_r - PL \cos q.
\end{equation}
By setting this second variation equal to zero, we find the critical buckling force
\begin{equation}
\label{eq-Pc}
P_c =\frac{k_r}{L \cos q},
\end{equation}
and by inserting equation~\ref{eq-Pq} into equation~\ref{eq-d2V} we see that this postbuckling path is stable for all $q$.  For $P \leq P_c$, the bar will remain vertical and undeflected, and for $P > P_c$, the bar will buckle either to the left or the right, breaking symmetry, and this buckling direction will be dictated by imperfections in the bar or the eccentricity of the load. 

\section{Buckling of continuous structures}

\subsection{Elastica}

The history behind the mathematical treatment of a thin elastic is rich, and a detailed discussion of it is well beyond the scope of this review. I would encourage all interested to read a history of it as detailed by Levien~\cite{Levien2008}. A nice recent article that revisited the mathematical intricacies of the problem was prepared by Singh {\em et al.}~\cite{Singh2017}, along with a detailed derivation of the governing equations and various solutions in a recent book by Bigoni~\cite{Bigoni2015}. There are typically three flavors of equilibrium equations for the elastica, and their utility very much depends on either what community the research is embedded within, or how the particulars of a given problem lend themselves to a straightforward solution. In general, you will encounter the equations parameterized by the angle along the arc length of the elastic curve $\theta(s)$, the curvature along the arc length $\kappa(s)$, or through a force and moment balance. All three are equivalent and can be recovered with a little effort and some geometry -- an exercise left up to the reader. For instance, a moment and force balance yields
\begin{subequations}
\label{eq-elastica-fm}
\begin{align}
\vett{m}'(s) +\hat{\vett{t}}(s) \times \vett{f}(s) &=0, \\
\vett{f}'(s)+\hat{\vett{n}}(s)&=0,
\end{align}
\end{subequations}
where $\vett{m}(s)$ and $\vett{f}(s)$ are the moments and forces acting on the curve, respectively, and $\hat{\vett{t}}(s)$ and $\hat{\vett{n}}(s)$ are the tangent and normal vectors along the curve, respectively. Here, we use an apostrophe to denote an ordinary derivative with respect to the arc length. Using the Frenet--Serret frame, the equations can be rewritten in terms of the curvature along the arc length as
\begin{equation}
\label{eq-elastica-k}
2\kappa''(s)+\kappa^3-\mu \kappa(s) +\sigma =0,
\end{equation}
where $\mu$ and $\sigma$ are constants corresponding to how the elastica is loaded. Finally, this equation can in turn be shown to be equivalent to the equilibrium equation of a planar elastica as parameterized by the angle along the arc length as
\begin{equation}
\label{eq-elastica}
\theta''(s)+\lambda_p^2 \sin \theta(s) =0,
\end{equation}
where $s$ is the arc length that parameterizes the curve, $\theta$ is the angle that the tangent vector at a given point $s$ makes with the horizon, and $\lambda_p^2=P/B$ is the ratio of the applied load $P$ to the bending rigidity $B$ of the beam. 

The buckling of an elastica, or an Euler column, is a problem encountered by most engineers during their studies. Its solution begins by linearizing equation~\ref{eq-elastica}, which is nonlinear because of the term $\sin \theta(s)$. Linearizing about the flat state allows us to consider small angles, such that $\sin \theta(s) \approx \theta(s)$, such that what started as a nonlinear eigenvalue problem now becomes a linear eigenvalue problem. For example, for an elastic that is simply supported at its ends, the lowest eigenvalue is gives a critical buckling load of
\begin{equation}
\label{euler}
P_c = \pi^2\frac{B}{L^2},
\end{equation}
where $L$ is the length of the elastica. For a clamped--clamped elastica, the buckling load is four times as large. Linearizing the equations comes at a cost, however --  while we gain insight into the force at which buckling will occur, along with the mode shape of the buckled structure, we lose the ability to quantify the amplitude of the deflection. 

\begin{figure} 
\begin{center} 
\includegraphics[trim=0 0 0 35, clip, width=1\columnwidth]{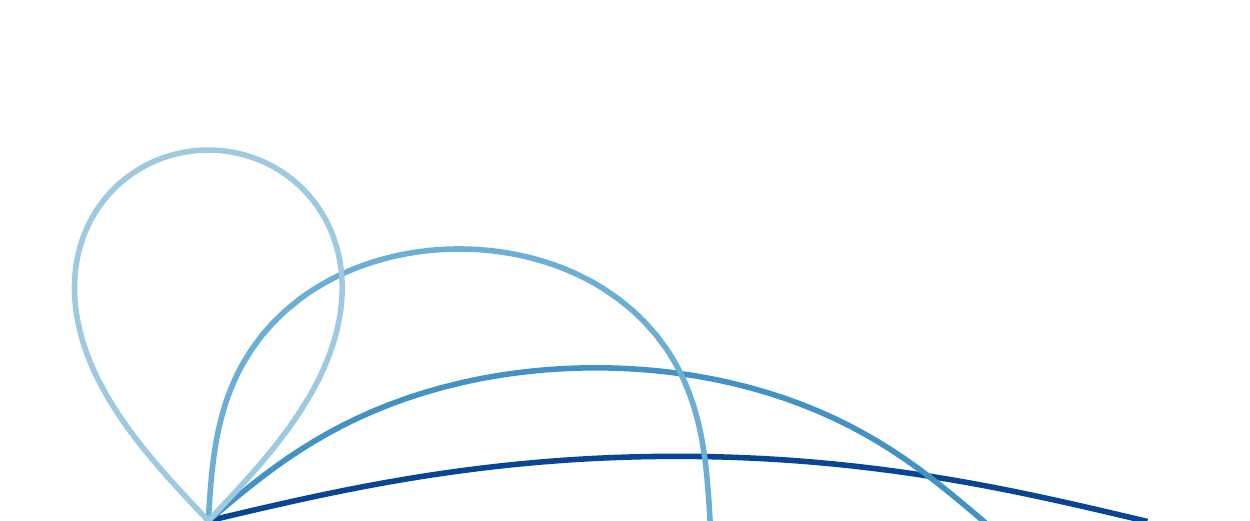} 
\includegraphics[trim=0 0 0 35, clip, width=1\columnwidth]{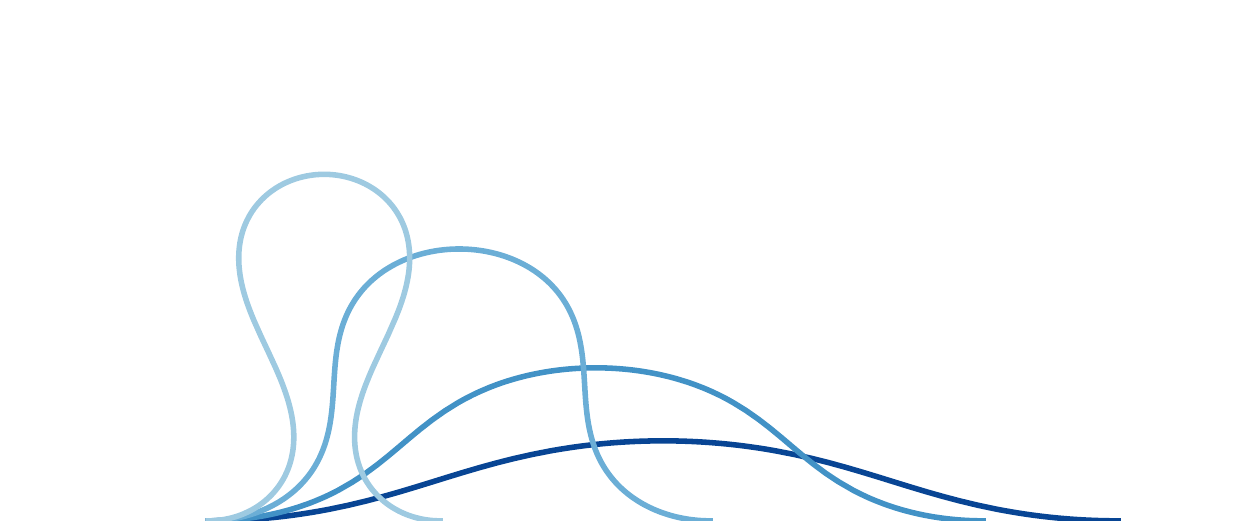} 
\end{center}
\vspace{-7mm} 
\caption{(Top) Shapes of a simply supported elastica subjected to various end shortenings. Graphs were generated in \textsc{Mathematica} using equations~\ref{eq-xs},b. (Bottom) Shapes of a clamped--clamped elastica subjected to various end shortenings. Graphs were generated in \textsc{Mathematica} using equations~\ref{eq-xc},b using the code block provided below these equations. \label{fig-elastica}} 
\vspace{-4mm} 
\end{figure}

The post--buckling shape of a simply supported elastica is determined by the parametric equations~\cite{Bigoni2015}
\begin{subequations}
\begin{align}
\label{eq-xs}
x(s) &= \frac{2}{\Lambda(k)}(\mathcal{E}[\text{am}[s \Lambda(k) +\mathcal{K}[k]|k]|k] \nonumber \\ &\hphantom{{}= \frac{2}{\Lambda(k)}}-\mathcal{E}[am[\mathcal{K}[k],k],k])-s, \\
y(s) &= -\frac{2}{\Lambda(k)}k\text{cn}[s\Lambda(k)+\mathcal{K}[k]|k],
\end{align}
\end{subequations}
where $k=\sin{\frac{\alpha}{2}}$, $\alpha$ is the angle of rotation at the inflection point at $s=L/2$, {\em i.e.} symmetry allows the analysis to focus on only half of the rod length, $\Lambda(k)=2m\mathcal{K}[k]$ which for a mode one deformation $m=1$, $\mathcal{K}[\cdot]$ is the complete elliptic integral of the first kind, $\mathcal{E}[\cdot|\cdot]$ is the incomplete elliptic integral of the second kind, $\text{am}[\cdot|\cdot]$ is the amplitude for Jacobi elliptic functions, and $\text{cn}[\cdot | \cdot]$ is the Jacobi cn elliptic function. Increasing $\alpha$ will increase the amplitude of the elastica while conserving the elastica's arc length. Similarly, the post--buckling shape of clamped--clamped elastica is then determined by the parametric equations~\cite{Bigoni2015}
\begin{subequations}
\begin{align}
\label{eq-xc}
x(s) &= \frac{2}{\Lambda(k)}\mathcal{E}[\text{am}[s \Lambda(k)|k]|k]-s, \\
y(s) &= \frac{2}{\Lambda(k)}k(1-\text{cn}[s\Lambda(k)|k]),
\end{align}
\end{subequations}
where $k=\sin{\frac{\alpha}{2}}$, $\alpha$ is the angle of rotation at the inflection point at $s=L/4$, {\em i.e.} symmetry allows the analysis to focus on only quarter of the rod length, and $\Lambda(k)=2(m+1)\mathcal{K}[k]$ which for a mode one deformation $m=1$. Increasing $\alpha$ will increase the amplitude of the elastica while conserving the elastica's arc length. The following \textsc{Mathematica} code is provided to enable the reader to visualize the deformed shape of the clamped--clamped elastica given by equations~\ref{eq-xc}
\begin{mmaCell}[]{Input}
k[\mmaUnd{\(\alpha\)}_]:= Sin[\mmaUnd{\(\alpha\)}/2]
\end{mmaCell}
\begin{mmaCell}{Input}
\mmaUnd{\(\lambda\)}c[k_, m_]:= 2 (m + 1) EllipticK[k] 
\end{mmaCell}
\begin{mmaCell}{Input}
xc[s_, \mmaUnd{\(\alpha\)}_, m_]:= (-s + 
   2/\mmaUnd{\(\lambda\)}c[k[\mmaUnd{\(\alpha\)}], 
     m] (EllipticE[JacobiAmplitude[
     s \mmaUnd{\(\lambda\)}c[k[\mmaUnd{\(\alpha\)}], m], k[\mmaUnd{\(\alpha\)}]], k[\mmaUnd{\(\alpha\)}]]))
\end{mmaCell}
\begin{mmaCell}{Input}
yc[s_, \mmaUnd{\(\alpha\)}_, m_]:= (2 k[\mmaUnd{\(\alpha\)}]/\mmaUnd{\(\lambda\)}c[k[\mmaUnd{\(\alpha\)}], 
    m] (1 - JacobiCN[s \mmaUnd{\(\lambda\)}c[k[\mmaUnd{\(\alpha\)}], m], 
    k[\mmaUnd{\(\alpha\)}]]))
\end{mmaCell}
\begin{mmaCell}{Input}
ParametricPlot[
\{xc[s, 1.5, 1], yc[s, 1.5, 1]\},
 \{s, 0, 1\}]
\end{mmaCell}
where for the parametric plot, values of $\alpha=1.5$ and mode number of $m=1$ were chosen arbitrarily. Example elastica curves are plotted in Fig.~\ref{fig-elastica}.

In both the simply supported and clamped--clamped case, the end shortening $u(\alpha)/L$ is given by
\begin{equation}
\frac{u(\alpha)}{L}=2 \left(1-\frac{\mathcal{E}(k)}{\mathcal{K}(k)}\right).
\end{equation}

\subsection{Plates \& Shells}

We began this review with Koiter's thin shell equations, {\em i.e.} equations~\ref{Us} and~\ref{Ub}. There are various simplifications that can be made to these equations to make them more mathematically tractable~\cite{Niordson1985}, and we will begin with a commonly used set of equations to describe the mechanics of thin plates: the F\"{o}ppl-von K\'{a}rm\'{a}n (FvK) plate equations. The FvK equations are rooted in an approximation for the plate's in--plane strain $\varepsilon_{\alpha \beta}$ where the non--linear terms describing the plate's out--of--plane deflection $w$ are retained, while the non--linear terms corresponding to the plate's in--plane displacements $u_{\alpha}$ are discarded, such that $\varepsilon_{\alpha \beta} \approx 1/2(u_{\alpha,\beta} + u_{\beta,\alpha}+ w_{,\alpha} w_{,\beta}$), where the comma represents differentiation, {\em i.e.} $f_{,\alpha} \equiv \partial f/\partial x_\alpha$. 
Using the FvK approximation for strain, a plate's equilibrium equations can be arrived at from a variational approach that minimizes the plate's free energy~\cite{Landau1959}.
\begin{subequations}
\begin{align}
\label{fvk1ME:dim}
B\nabla^4w -h \sigma_{\alpha\beta}w_{,\alpha\beta}&=0, \\
\sigma_{\alpha\beta,\beta}&=0,
\label{fvk2ME:dim}
\end{align}
\end{subequations} 
where $B=Eh^3/12(1-\nu^2)$ is the bending stiffness. Alternatively, by introduction of the Airy potential, $\sigma_{\alpha\beta}=\epsilon_{\alpha\gamma}\epsilon_{\beta\mu}\phi_{,\lambda\mu}$, where $\epsilon_{\alpha\beta}$ is the two--dimensional Levi--Civita symbol, and the ``die'' operator, which represents a symmetric contraction of two components,{\em i.e.} $\lozenge^4[f,g]\equiv f_{,\alpha\alpha}g_{,\beta\beta}-2f_{,\alpha\beta}g_{,\alpha\beta}+f_{,\beta\beta}g_{,\alpha\alpha}$~\cite{Mansfield1989}, the equilibrium equations can be written as
\begin{subequations}
\begin{align}
\label{fvk1:dim}
B\nabla^4w -\lozenge^4[\phi,w]&=0, \\
\frac{1}{Y}\nabla^4\phi + \frac{1}{2}\lozenge^4[w,w]&=0,
\label{fvk2:dim}
\end{align}
\end{subequations} 
where $Y=Eh$ is the stretching stiffness. These equations are non--linear, with the term $\lozenge^4[\phi,w]$ coupling the plate's curvature with in--plane stress, and the term $\lozenge^4[w,w]/2$ representing the Gauss curvature $K$. The explicit appearance of the Gauss curvature illustrates the intimate connection between a plate's elasticity and its geometry~\cite{Audoly2010}, and this interplay will consistently appear in the various examples that follow. 

A plate is thin structure that is flat in its stress--free configuration, while a shell has a non--zero initial curvature. An instructive approach to the complexities that arise with shell mechanics is to regard the shell as consisting of two distinct surfaces: one which sustains {\em stretching} stress resultants and the other which sustains the {\em bending} stress resultants. Calladine~\cite{calladine} covers this {\em two--surface} approach to the equilibrium equations for shells in a comprehensive manner, and we will only highlight the key components here. In this two--surface approach, the stretching surface is equivalent to the membrane hypothesis of shells, and the bending surface is very similar to the FvK analysis detailed above. The general interaction between these two imaginary surfaces, {\em i.e.} the way an externally applied force is distributed between them, leads directly to the Donnell--Mushtari--Vlasov (DMV) equations~\cite{soedel04}. 
\begin{subequations}
\begin{align}
\label{dmv1:dim}
B\nabla^4w+\nabla_k^2\phi -\lozenge^4[\phi,w]&=0, \\
\frac{1}{Y}\nabla^4\phi -\nabla_k^2 w +\frac{1}{2}\lozenge^4[w,w]&=0.
\label{dmv2:dim}
\end{align}
\end{subequations} 
These equilibrium equations contain the Vlasov operator $\nabla_k^2(f)\equiv R_\beta^{-1}f_{,\alpha\alpha}+R_\alpha^{-1}f_{,\beta\beta}$, which incorporates the shell's principal curvatures. It should be immediately apparent that in the absence of any initial curvature, equations~\ref{dmv1:dim} and \ref{dmv2:dim} revert directly to the FvK equations given by equations~\ref{fvk1:dim} and \ref{fvk2:dim}. Since these equations are an extension of the FvK plate equations, they retain the same variational structure, and are applicable under similar assumptions of small strains and moderate rotations~\cite{Audoly2010}.

Few exact solutions to the F\"{o}ppl-von K\'{a}rm\'{a}n equations are known to exist, but geometrical simplifications make the FvK equations useful for describing the Euler buckling of a plate in response to an in--plane stress. A uniaxially compressed plate that buckles out--of--plane will have cylindrical symmetry, which greatly simplifies the geometry, since a cylindrical shape is developable to a plane, such that in equation~\ref{fvk2:dim} $K=\lozenge^4[w,w]/2=0$. Since there is only stress in one direction, say the $x$--direction, equation~\ref{fvk2:dim} reduces to $\phi_{,xxxx}=0$. Considering a plate with clamped boundary conditions, this equation for $\phi$ can be integrated, allowing equation~\ref{fvk1:dim} to reduce to an ordinary differential equation, which can be solved in terms of the plate's deflection~\cite{Audoly2010},
\begin{equation}
\label{fvk-buckle}
w(x,y)=\pm \frac{h}{\sqrt{3}}\left(\frac{\sigma}{\sigma_{Eu}}-1\right)^{1/2}\left(1+\cos{\frac{\pi x}{L}}\right),
\end{equation}
where $L$ is the length of the plate, $\sigma$ is the uniaxial stress, and the Euler buckling stress is
\begin{equation}
\sigma_{Eu}=\frac{\pi^2E}{3(1-\nu^2)}\left(\frac{h}{L}\right)^2.
\end{equation}
As noted in the subsection on {\em Warping Wafers}, buckling of thin plates can also occur from a mismatch of strains through the thickness, in particular with heated or differentially swollen plates. Some additional, relevant references to this topic include~\cite{Mansfield1962, Mansfield1965,Masters1993, Salamon1995, Freund2000, Seffen2007}.

\subsection{Wrinkling}

When a thin plate is bound to a compliant substrate and compressed, higher buckling modes emerge, and the pattern formation of these ordered buckled structures, or wrinkles, have garnered significant recent interest. The wavelength of wrinkles is selected by a balance of plate's bending energy and the energy required to deform the underlying elastic substrate. The bending resistance of the sheet penalizes short wavelengths, while deformation of the elastic foundation that is supporting the sheet penalizes long wavelengths. An intermediate wavelength emerges when we consider that the reaction of the underlying layer $K$ is proportional to the deflection of the plate $w$. In the simplest case, for 1D wrinkles extending in the $x$--direction, equation~\ref{fvk1ME:dim} becomes
\begin{equation}
\label{fvk1ME:dim-1Dwrinkle}
Bw_{,xxxx}-h\sigma_{xx}w_{,xx}+Kw=0,
\end{equation}
when a Winkler foundation is included~\cite{Dillard2018}. By linearizing this equation, and discarding any stretching of the mid--plane due to curvature, {\em i.e.} the second term is zero, a characteristic length scale emerges based on a balance of the two rigidities. The scaling we found in equation~\ref{wrinkleLam} can be recovered here from dimensional analysis. In the limit of wrinkles on a very deep substrate, {\em i.e.} $h \ll \lambda \ll H_s$, the stiffness of the elastic foundation scales as $K \sim E_s/\lambda$~\cite{Cerda2003}, and the wrinkle wavelength therefore scales as
\begin{equation}
\label{lambda_wrinkle}
\lambda \sim h\left(\frac{E}{E_s}\right)^{1/3}.
\end{equation}
We can explicitly determine the critical stress required for wrinkles to form and the resulting wavelength by linearizing equations~\ref{fvk1ME:dim} and~\ref{fvk2ME:dim} about the flat, unbuckled state, {\em i.e.} $u_\alpha=w=0$, and performing linear stability analysis~\cite{Allen1969,Chen2004a,Chen2004b,Audoly2008a,Audoly2008b,Audoly2008c, Cai2011}. Linearizing these equations requires the strain tensor to simplify to $\varepsilon_{\alpha \beta} \approx 1/2(u_{\alpha,\beta} + u_{\beta,\alpha})$. The plate will be considered infinite in the $x_\alpha$ directions, and exposed to equibiaxial stress, such that the linearized equations are $B\nabla^4 w-h \sigma_0 \nabla^2w=-p$ and $\nabla^4 \phi=0$. The stress component exerted by the substrate onto the plate is introduced as $p$. These ordinary differential equations allow periodic solutions in the form $w(x,y)=\hat{w}\cos{(k_1x)}$, which result in a cylindrical pattern described by a critical threshold $\sigma_c$ and wavelength $\lambda$ given by~\cite{Groenewold2001, Chen2004a, Chen2004b, Audoly2008b}
\begin{subequations}
\begin{align}
\label{wrinkle_c}
\sigma_c&=E^*\left(\frac{3E_s^*}{2E^*}\right)^{2/3},  \\
\label{wrinkle_lam}
\lambda &= \pi h\left(\frac{2E^*}{3 E_s^*}\right)^{1/3}.
\end{align}
\end{subequations}
The starred quantities $E^*$ and $E_s^*$ represent the effective, or reduced, modulus of the plate and substrate, respectively. The reduced modulus of the plate is $E^*=E[1-\nu^2]^{-1}$, while the reduced modulus of the substrate includes the tangential traction forces exerted by the substrate onto the plate, such that $E_s^*=E_s(1-\nu_s)[(1+\nu_s)(3-4\nu_s)]^{-1}$~\cite{Audoly2008a, Audoly2008b, Audoly2008c}. 

Wrinkles that form under compression are familiar to most everyone -- they appear by simply compressing our skin. It is less intuitive to observe wrinkles that form as a free elastic sheet is pulled in tension~\cite{Friedl2000, Cerda2002, Cerda2003, Nayyar2011}. This problem is no longer confined to 1D as there is a tension $T$ in the $y$--direction, and localized regions of compressive in the $x$--direction near the clamped boundaries. The equilibrium equation that emerges from minimizing the free energy for this configuration is $Bw_{,xxxx}-h\sigma_{xx}w_{,xx}-Tw_{,yy}=0$~\cite{Cerda2003}. Cerda {\em et al.}~\cite{Cerda2003} identified an analogy between the energy in an elastic foundation supporting a thin sheet, $U_f \sim \int_A \! Kw^2 \ \text{d} A$,and the sheet's stretching energy, $U_m\! \sim  \! \int_A \! T\left(w_{,x}\right)^2 \ \text{d} A$. Since these energies are of similar form, and with the in--plane strain scaling as $w_{,x}\sim w/L$, comparing the two leads to the emergence of {\em effective} stiffness of the elastic foundation, $K\sim T/L^2$.  This connection is convenient as it allows equation~\ref{fvk1ME:dim-1Dwrinkle}, and the scaling in equation~\ref{lambda_wrinkle}, to hold for a variety of wrinkling problems, with each problem varying only in the actual form of the effective stiffness of the elastic foundation, $K$. 

\subsection{Folding}

As the amount of overstress, or confinement, of the compressed plate increases, the bending energy along the plate goes from being broadly distributed among wrinkles to being localized within sharper features~\cite{Huang2007, Pocivavsek2008, Holmes2010, Brau2010, Huang2010, Vella2010b, Davidovitch2011, Davidovitch2012, King2012, Ebata2012, Schroll2013, Diamant2013, Vella2015, Paulsen2016, Box2017, Vella2018}. When the compressive strain exceeds a critical value of $\varepsilon \simeq 0.2$, a pitchfork bifurcation of the wrinkle morphology emerges as one wrinkle grows in amplitude while neighboring wrinkles decrease~\cite{Pocivavsek2008, Holmes2010, Brau2010}. This focusing of bending energy leads to a break in the up--down symmetry of the wrinkled plate, and is analogous to period--doubling bifurcations in dynamical systems~\cite{Brau2010}. The symmetry breaking occurs from the nonlinear contribution of the compliant foundation that supports the stiff plate as the out--of--plane deflection and in--plane compression of the plate are no longer equivalent~\cite{Brau2010, Diamant2013}. Further compression of the plate beyond a strain of $\varepsilon \simeq 0.26$ leads to a period-quadrupling bifurcation~\cite{Brau2010}, in which sharp folds appear, localizing the much of the stress within highly curved ridges~\cite{Witten2007, Pocivavsek2008, Holmes2010, Brau2010, Cambou2011, Diamant2013}. This folding is a means for focusing the elastic energy within the plate.  Qualitatively, a fold occurs when the radius of curvature of the deformed feature is on the same order as the thickness of the film.

When folds occur over a large scale, for instance with the crumpling of a piece of paper, these stress-focused ridges can significantly alter the mechanical properties of the structure. A crumpled piece of paper is characterized by sharp ridges that terminate at point--like singularities. These singularities are conical dislocations, and they emerge as the sheet tries to resist stretching, thereby localizing the stretched region into the tip of the cone. Referred to as {\em developable cones}, or {\em d--cones}, due to their isometry to a flat plate, the shape is a particular solution to the FvK equations in the limit of large deflections~\cite{BenAmar1997, Cerda1998, Chaieb1998, Cerda1999, Chaieb2000}. A d--cone is a building block to a crumpled sheet, and two of these singularities are connected by a stretching ridge, which contains much of the deformation energy within a strongly buckled sheet~\cite{Witten1993, Lobkovsky1995, Lobkovsky1997}. The width of these stretching ridges can be arrived at either by scaling considerations of the sheets elastic energy~\cite{Lobkovsky1995} or by a boundary layer analysis of the FvK equations~\cite{Lobkovsky1996}. Upon formation of a ridge, the bending and stretching energies are of the same order, and for a sharp fold on a plate with a given thickness, these energies will scale linearly with the crease length $\chi$, {\em i.e.} $U_m\sim U_b\sim \chi$. In reality, the curvature of the fold is softened by a small amount of stretching along the ridge, and as the length of the fold increases these energies grow at a slower rate, $U_m\sim U_b\sim \chi^{1/3}$~\cite{Witten1993, Lobkovsky1995, Lobkovsky1996, Lobkovsky1997}. Recent work has focused on the dynamic deformations of d--cones and ridges~\cite{Hamm2004, Boudaoud2007}, the existence and annihilation of multiple singularities in a constrained elastic plate to minimize stretching~\cite{Boudaoud2000, Walsh2011}, the mechanics of curved--folds~\cite{Dias2012}, and the nature of dipoles in thin sheets~\cite{Guven2013}.

\end{document}